\documentclass[journal]{IEEEtran}
\usepackage{amsmath,amssymb}
\usepackage{subfigure}
\usepackage{bm}
\usepackage{graphicx,graphics,psfrag}
\usepackage{cite,balance}
\usepackage{caption}
\allowdisplaybreaks
\usepackage{algorithm}
\usepackage{algorithmic}
\usepackage{accents}
\usepackage{amsthm}
\usepackage{url}
\usepackage[english]{babel}
\usepackage{multirow}
\usepackage{enumerate}
\usepackage{cases}
\usepackage{stfloats}
\usepackage{dsfont}
\usepackage{color,soul}
\usepackage{amsfonts}
\usepackage{fancyhdr}
\usepackage{hhline}
\usepackage{array}
\usepackage{mathtools}

\graphicspath{{./Figs/}}

\newtheorem{theorem}{\emph{\underline{Theorem}}}

\newtheorem{case}[theorem]{\emph{\underline{Case}}}

\newtheorem{lemma}{\emph{\underline{Lemma}}}

\newtheorem{proposition}{\emph{\underline{Proposition}}}

\newtheorem{remark}{\bf \emph{\underline{Remark}}}

\def\({\left(}
\def\){\right)}

\def\b0{{\mathbf{0}}}

\newcommand{\diag}{\mathrm{diag}}

\begin{document}
	\captionsetup[figure]{name={Fig.}} 
	\title{Rotatable IRS-Assisted 6DMA Communications:\\ A Two-timescale Design}
	\author{Chao Zhou, 
		Changsheng~You,~\IEEEmembership{Member,~IEEE}, 
		Cong Zhou,  
		Liujia Yao,
		Weijie Yuan,~\IEEEmembership{Senior Member, IEEE},\\
		Beixiong Zheng,~\IEEEmembership{Senior Member, IEEE},
		and Nan Wu,~\IEEEmembership{Senior Member, IEEE}
		\thanks{Chao Zhou, Changsheng You, Cong Zhou, and Liujia Yao are with the Department of Electronic and Electrical Engineering, Southern University of Science and Technology (SUSTech), Shenzhen 518055, China (e-mail:~zhouchao2024@mail.sustech.edu.cn, youcs@sustech.edu.cn, zhoucong@stu.hit.edu.cn, yaolj2024@mail.sustech.edu.cn). 
				
		Weijie Yuan is with the School of System Design and Intelligent Manufacturing and the Shenzhen Key Laboratory of Robotics and Computer Vision, Southern University of Science and Technology, Shenzhen 518055, China (e-mail: yuanwj@sustech.edu.cn)	
		
		Beixiong Zheng is with the School of Microelectronics, South China University of Technology, Guangzhou, China (e-mail: bxzheng@scut.edu.cn).
		
		Nan Wu is with the School of Information and Electronics, Beijing Institute of Technology, Beijing 100081, China (e-mail: wunan@bit.edu.cn).
		
		\emph{(Corresponding author: Changsheng You.)}  
		}\vspace{-14pt}} 
	
	\maketitle
	\begin{abstract}
		Intelligent reflecting surface (IRS) and movable antenna (MA) have emerged as two promising technologies to improve wireless communication performance by proactively reconfiguring wireless channels at the environment and transceiver sides, respectively. However, the performance of both IRS and MA systems is also constrained by practical limitations. 
		To address this issue, we propose in this paper a new multi-functional antenna/surface communication system by exploiting their complementary advantages, where a \emph{rotatable} IRS (R-IRS) is deployed to enhance the downlink communications from a six-dimensional MA (6DMA)-equipped base station (BS) to multiple single-antenna users. 
		To reduce the prohibitively high complexity in real-time channel estimation and beamforming design, we formulate an optimization problem to maximize the average sum-rate based on a  two-timescale (TTS) transmission protocol. Specifically, the antenna configuration of BS (including both antenna position and rotation) as well as the rotation and reflection of IRS are optimized based on statistical channel state information (S-CSI), while only the transmit beamforming of BS is designed according to instantaneous CSI (I-CSI) in the short-timescale. 
		To obtain useful insights, we first consider a single-user case and show that the 6DMA in the BS should form a sparse array to achieve multi-beam transmissions towards both the IRS and user, hence allowing for efficient coordination of the direct and reflected channels, while the rotation of IRS is exploited to achieve effective multi-path alignment.
		Then, for the general multi-user case, the optimization problem is non-convex and challenging to solve. To tackle this difficulty, we propose an efficient algorithm to obtain a high-quality solution
		by using the classic weighted minimum mean-square error (WMMSE) and stochastic successive convex approximation (SSCA) techniques. Moreover, a low-complexity algorithm is further proposed to 
		reduce the computational complexity.
	 	Numerical results validate the effectiveness of proposed multi-functional antenna/surface system over various benchmarks and highlight the performance gains achieved by jointly exploiting the spatial degrees-of-freedom of the 6DMA-BS and R-IRS under the TTS transmission protocol.	
	\end{abstract}
	\vspace{-10pt}
	\vspace{-10pt}
	\section{Introduction} 
	The future sixth-generation (6G) wireless systems demand more stringent performance requirements than the fifth-generation (5G), such as ubiquitous connectivity, extremely high data rates, and low latency~\cite{6G_road,YouNGAT}. However, these requirements may not be fully achieved by conventional multiple-input multiple-output (MIMO) systems that mainly adapt to random and uncontrollable wireless channels, but without the capability to alter the wireless channel itself. This thus motivates two efficient technologies in recent years, namely, \emph{intelligent reflecting surface} (IRS)~\cite{wu2024intelligent} and \emph{movable antenna} (MA)~\cite{ZhuMAtutorial}, that proactively reconfigure wireless channels at the environment and transceiver sides, respectively. 
	However, the performance of IRS-only systems may be constrained by the limited spatial degrees-of-freedom (DoFs) at the transceiver side due to fixed-position/rotation antenna configuration, while MA-only systems may suffer unsatisfactory communication performance if there exist large-sized blockages in the communication links.	
		
	To address the above issues, we propose in this paper
	 a new multi-functional array/surface wireless system by synthesizing the advantages of both IRS and MA. Specifically, a \emph{rotatable IRS (R-IRS)} is deployed to assist a \emph{six-dimensional MA (6DMA)}-equipped base station (BS) in serving multiple single-antenna users. 
	We show that the sparse configuration of MA enables multi-beam generation, thereby facilitating efficient coordination of the direct and reflected channels. Additionally, the rotation of the IRS enables flexible multi-path alignment with low implementation complexity. Consequently, this joint design of both the MA and IRS contributes to enhanced rate performance.
	\vspace{-10pt}
	\subsection{Related Works}
	\subsubsection{Intelligent Reflecting Surfaces}
	IRSs (also known as reconfigurable intelligent surfaces (RISs) that share similar signal control principle) are a type of low-cost metasurfaces that can smartly control the amplitudes and/or phase-shifts of reflected signals via a large number of passive reflecting elements~\cite{wu2024intelligent}. This property thus allows IRSs to enable a variety of new functions, such as establishing virtual line-of-sight links between transceivers to bypass environmental blockage, enhancing received signal power and mitigating multi-user interference, as well as transforming fast-fading channels into slow-fading ones for improving communication reliability~\cite{IRSCoverage,MyTWC}.
	To achieve the performance gains of IRSs, various passive beamforming designs and channel estimation methods have been proposed in the literature (see, e.g.,~\cite{ZhengIRSCE_survey} and references therein). Specifically, for the passive beamforming design, efficient techniques, such as successive convex approximation (SCA) and semi-definite relaxation (SDR), were proposed to optimize IRS reflection coefficients under the unit-modular constraint~\cite{ZhengIRSCE_survey}, which, however, may become computationally forbidding when the number of reflecting elements is very huge. On the other hand, IRS beamforming design entails the acquisition of accurate channel state information (CSI), which is practically challenging due to its passive reflection nature as well as the large number of reflecting elements. This thus motivates a line of research that proposed various methods to reduce the number of pilots for estimating the cascaded IRS channels~\cite{DaiIRSCE,LiuIRSCE}. Apart from \emph{instantaneous} CSI (I-CSI) acquisition and beamforming design, an alternative approach is by using the two-timescale (TTS) protocol, where \emph{statistical} CSI (S-CSI) of IRS channels is estimated to design efficient IRS reflections in the long-timescale, while the active beamforming of BS is designed in the short-timescale based on instantaneous and low-dimensional CSI~\cite{Zhaomm2021TSS}.
	However, the performance of IRS-only systems may be fundamentally constrained by the product-distance path-loss that necessitates a huge number of reflecting elements to compensate for the severe path-loss~\cite{MyTWC}.  Moreover, IRSs can flexibly control the reflected links, while they cannot dynamically reconfigure the direct BS-user channels and thus lack the DoF to fully control/coordinate the communication channels, including both reflected and direct links.

	\subsubsection{Movable Antennas}
	MAs are another promising technology to reconfigure wireless channels by adjusting the antenna positions at the transmitter and/or receiver. By exploiting the new spatial DoFs in antenna configuration, MAs enable flexible control of wireless channels in the field response, hence achieving appealing advantages over fixed-position array (FPA) systems in terms of interference mitigation, signal enhancement, spatial multiplexing, etc.~\cite{ZhuMAtutorial}. Specifically, the authors in~\cite{Zhu2024MA_performance} and~\cite{Ma2024_MAMIMO} demonstrated that optimizing MA positions enables efficient multi-path phase alignment, thereby enhancing channel capacities of both single-input single-output (SISO) and MIMO systems. Moreover, the positionable DoFs offered by MAs also facilitate efficient beam management.
	Building on these advantages, several studies have exploited MAs to improve physical-layer security by employing, e.g., gradient-based optimization~\cite{Hu2024MA_Secure} and heuristic algorithms~\cite{MAcovert}. In addition, by flexibly adjusting antenna positions, MAs can realize various array configurations for enabling a broad range of applications such as wireless sensing~\cite{MAsensing} and spectrum sharing~\cite{Zhou_MA}.
	Beyond MAs, the recently proposed 6DMAs~\cite{Shao20256DMA} offer enhanced flexibility in optimizing wireless propagation environment by jointly exploiting both positionable and rotatable DoFs~\cite{Shao6DMAtutorial,Shao6DMAJSAC}. As a special case, RAs provide a simple yet cost-effective approach for reconfiguring wireless channels by controlling antenna orientation only, which has also garnered significant attention~\cite{zheng2025rotatable,ZhengRA,MyRA}.
	Although several studies have investigated channel estimation for MA and 6DMA systems~\cite{XiaoMACE,Shao6DMAtutorial}, obtaining real-time CSI in fast-fading scenarios remains challenging. Moreover, even with perfect I-CSI, real-time antenna position adjustment results in a prohibitively high hardware cost and implementation complexity. To address this issue, a TTS transmission protocol was adopted in~\cite{MATTS}, which effectively reduces the channel estimation overhead and computational complexity while achieving notable performance gains over conventional FPA systems.  
	Despite these advantages, the performance gain of MAs in improving channel capacity is particularly notable in rich-scattering environments, which diminishes under weak multi-path conditions. Moreover, in scenarios where the MA–user channels are severely blocked, the deployment of MAs becomes ineffective due to the restricted array aperture size in practice.
	\subsection{Motivations and Contributions}
\begin{figure*}[t]
 		\centering
 		\includegraphics[width=0.7\textwidth]{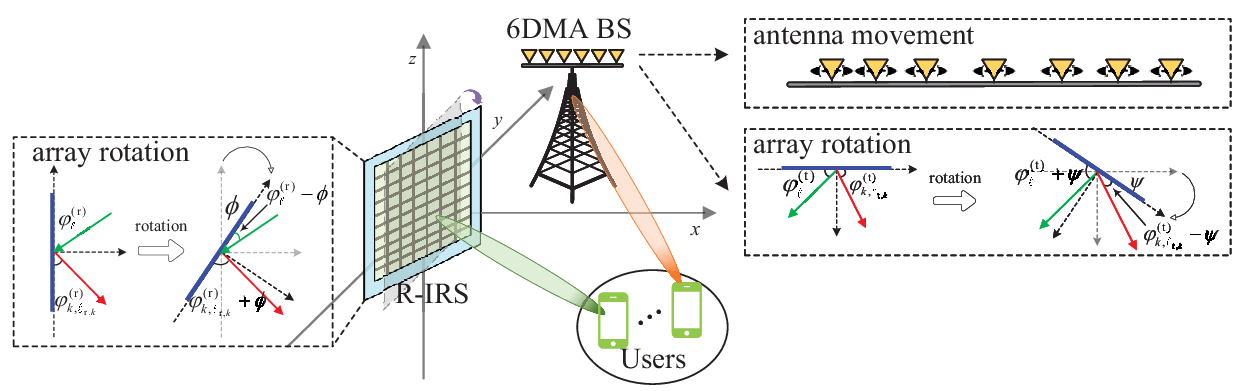}
 		\caption{The considered multi-functional antenna/surface-assisted multi-user communication systems.} \label{Fig:systemmodel}
 		\vspace{-14pt}
 	\end{figure*}

	Motivated by the above, we propose in this paper a multi-functional antenna/surface wireless system, as shown in Fig.~\ref{Fig:systemmodel}. By leveraging the positionable and rotatable DoFs of 6DMA-BS, together with rotation and phase shift adjustment of R-IRS, the equivalent BS-user channel (comprising both direct and reflected links) can be flexibly reshaped to achieve favorable phase alignment, improved interference suppression, as well as enhanced coverage performance, compared with existing IRS-only and MA-only systems~\cite{NingMAIRS,WuMAIRS,PanMAIRS,MAIRS2025}. 
	Moreover, it is worth noting that although some recent works~\cite{MSTAR_RIS,geng2024joint} have considered MA-IRS that allows IRS elements to move within a confined region, it usually requires additional active hardware architectures (e.g., mechanical motors) to enable reflecting element movement, hence incurring considerably high hardware cost and complexity in practice. In contrast, the proposed R-IRS only needs a rotatable shaft, hence offering a more cost-effective means to endow the new rotatable DoFs for achieving efficient multi-user interference mitigation.
	
	Despite these promising advantages, the new multi-functional wireless system also introduces several design challenges in practice. 
	First, acquiring accurate I-CSI of 6DMA and R-IRS-related channels is challenging, especially in rapidly time-varying environments. This challenge is further exacerbated by the
	strong reliance of I-CSI on the specific antenna and surface configurations, rendering it impractical to obtain I-CSI for all possible configurations.	 
	Second, adjusting the antenna configuration of BS (i.e., antenna position and rotation) as well as the rotation and reflection coefficients of IRS in real time generally incurs demanding implementation complexity, which becomes even unaffordable in fast-fading scenarios. 
	Third, the joint optimization of 6DMA-BS and R-IRS involves a large number of highly coupled parameters, making the problem non-convex and challenging to solve.
	Last but not least, the performance gains achieved by integrating 6DMA-BS and R-IRS in multi-user systems remain unclear, which deserves in-depth investigation.
	
	To address these issues,  we adopt an efficient \emph{TTS transmission protocol} for the multi-functional antenna/surface system. Specifically, the transmit beamforming of BS is designed based on I-CSI in the short-timescale, while the remaining variables (i.e., 6DMA position and rotation, IRS rotation, and reflection coefficients) are all optimized based on S-CSI in the long-timescale, hence greatly reducing the channel estimation overhead and implementation complexity.  To study the performance gain of proposed multi-functional antenna/surface-assisted multi-user communication systems, we formulate an optimization problem to maximize the average sum-rate of all users by jointly optimizing the long-timescale variables and the short-timescale beamforming vectors.
 	The main contributions of this paper are summarized as follows.
	\begin{itemize}
		\item 
		First, we consider a single-user case to gain fundamental insights into the proposed system. By designing the optimal transmit beamforming based on I-CSI in each short-timescale, the average rate maximization problem can be equivalently reformulated as the maximization of expected equivalent channel power gain. 
		Then, we show that arranging the 6DMA antenna elements in a sparse-array configuration facilitates favorable phase alignment between the direct and reflected links, thus enhancing the overall achievable rate. Additionally, the R-IRS exploits its rotatable DoFs to achieve multi-path alignment, further improving performance in multi-path scenarios. To solve this reformulated problem, we further decompose it into two subproblems: (i) optimizing the 6DMA positions and rotation, and (ii) optimizing the R-IRS rotation and reflection coefficients, which are efficiently solved in parallel by using the differential evolution (DE) algorithm and SDR technique, respectively.
		
		\item
		Next, for the multi-user case, we first employ the weighted minimum mean-square error (WMMSE) method to address the short-timescale beamforming design. Since there lacks a closed-form solution for the transmit beamforming vectors, the long-timescale optimization becomes more challenging. To tackle this difficulty, we reformulate the original problem into a two-layer structure: an inner problem for IRS reflection coefficient optimization, which is solved by using  stochastic successive convex approximation (SSCA) technique, as well as an outer problem for optimizing 6DMA position, rotation, and IRS rotation, which is solved by using extended DE algorithm.
		Furthermore, we develop a low-complexity algorithm in which the IRS reflection coefficients are optimized to maximize the average sum-channel gain, thereby avoiding iterative optimization required in the SSCA method and hence significantly reducing computational complexity.

		\item 
		Last, numerical results are presented to validate the effectiveness of proposed algorithms and the superiority of multi-functional antenna/surface in improving the rate performance under the TTS transmission protocol. It is shown that R-IRS rotation provides additional DoFs for enhancing reflected channel gain, while 6DMA position adjusting enables efficient phase alignment between directed and reflected channels, both leading to improved rate performance in the single-user case. In addition, benefiting from the flexible pattern design of 6DMA-BS and R-IRS, this integration
		effectively mitigates multi-user interference, showcasing a significant performance gain.
	\end{itemize}
	
	\section{System Model and Problem Formulation}\label{Sec2:label}	
	In this paper, we consider an R-IRS assisted multi-user communication system as shown in Fig.~\ref{Fig:systemmodel}, where a 6DMA-BS equipped with an $M$-antenna (denoted by $\mathcal{M}\triangleq\{1,2\ldots,M\}$) linear array (LA) communicates with $K$ (denoted by $\mathcal{K}\triangleq\{1,2\ldots,K\}$) single-antenna users, assisted by an R-IRS consisting of $N$ (denoted by $\mathcal{N}\triangleq \{1,2,\ldots,N\}$) reflecting elements in a planar array.
	Unlike conventional IRS-assisted communication systems where both BS antennas and IRS elements are at fixed location, the proposed system employs a 6DMA-BS to provide additional spatial DoFs at the transmitter side, as well as an R-IRS to endow the array rotation control for enabling adaptive communication coverage and flexible beam pattern design.
	Without loss of generality, we assume that the 6DMA-BS, the R-IRS center, and the users are located in the $x$-$y$ plane. In addition, the R-IRS is capable of dynamically adjusting its azimuth angle, while its elevation angle is fixed at zero.\footnote{In this work, we focus on a 1D rotation-and-movement array and a 1D R-IRS system, while more general scenarios involving multi-dimensional movable and rotatable antennas and/or IRS are left for future work.}
	For the linear 6DMA array, the position vector is denoted as $\mathbf{q} = \left[ q_1,q_2,\ldots,q_{M} \right]^T \in \mathbb{R}^{M \times 1}$ with $q_{m}$ representing the distance between the $m$-th antenna and 6DMA-BS center.
	\subsection{Channel Model}
	We consider the case where the users are located in the far-field region of both R-IRS and 6DMA-BS. In high-frequency band scenarios, the channel from the BS to the IRS can be modeled by the field-response model
	\begin{align}
		\mathbf{G}( \mathbf{q},\boldsymbol{\zeta}) \!=\!  \sum_{\ell = 0}^{L} \beta_{\ell} \mathbf{a}_{{\rm r},\ell} (\phi)  \mathbf{a}_{{\rm t},\ell}^H( \mathbf{q},\psi) 
		\!=\!\mathbf{G}_{\rm r}(\phi) 
		\mathbf{\Sigma} 
		\mathbf{G}_{\rm t}^{H}(\mathbf{q},\psi),
	\end{align}
	where $\boldsymbol{\zeta} = [\psi,\phi]^T$ with $\psi$ and $\phi$ denoting the 6DMA and IRS rotation angles, respectively. Additionally, $L$ denotes the number of scatterers with $ \ell = 0 $ and $ \ell \neq 0 $ representing the line-of-sight (LoS) path and non-LoS (NLoS) path, respectively. 
	In addition, $ \mathbf{G}_{\rm t}(\mathbf{q},\psi) \in \mathbb{C}^{ M \times (L+1) }$ denotes the field-response matrix (FRM) for the transmit region of BS, $ \mathbf{G}_{\rm r} (\phi) \in \mathbb{C}^{ N \times (L+1) } $ is the FRM for the receive region of IRS, and  $\mathbf{\Sigma} \in \mathbb{C}^{ (L+1) \times (L+1)}$ is the path-response matrix (PRM) from the BS to the IRS. Specifically, the transmit FRM consisting of $L+1$ field-response vectors (FRVs) can be expressed as
	\begin{align}\label{FRM_transmit}
		\mathbf{G}_{\rm t}(\mathbf{q},\psi ) =
		\big[
		\mathbf{a}_{{\rm t},0}(\mathbf{q},\psi),
		\ldots,
		\mathbf{a}_{{\rm t},L}(\mathbf{q},\psi)
		\big],
	\end{align}
	where $ \mathbf{a}_{{\rm t},\ell}( \mathbf{q},\psi) \in \mathbb{C}^{M\times 1} $ is given by
	\begin{align}\label{atk}
		\mathbf{a}_{{\rm t},\ell}(\mathbf{q},\psi)\! \triangleq\! 
		\Big[e^{\jmath \frac{2\pi}{\lambda}q_{1} \cos ( \varphi_{\ell}^{(\rm t)} \!+ \psi ) },\! 
		\ldots,\!
		e^{\jmath\frac{2\pi}{\lambda}q_{M} \cos( \varphi_{\ell}^{(\rm t)}\! + \psi  ) }
		\Big]^T,
	\end{align}
	with $ \varphi_{\ell}^{(\rm t)} $ denoting the angle-of-departure (AoD) of the $\ell$-th path and $\lambda$ representing the carrier wavelength. Similarly, the receive FRM of R-IRS consists of $ L +1 $ FRVs, which can be expressed as
	\begin{align}\label{FRM_receive}
		\mathbf{G}_{{\rm r}}(\phi )  = 
		\big[
		\mathbf{a}_{{\rm r},0} (\phi ),
		\mathbf{a}_{{\rm r},1} (\phi ),
		\ldots,
		\mathbf{a}_{{\rm r},L} (\phi )
		\big], 
	\end{align}
	Herein, the corresponding FRV $ \mathbf{a}_{{\rm r},\ell}(\phi) \in \mathbb{C}^{N\times 1} $ is given by
	\begin{align}
		\mathbf{a}_{{\rm r},\ell} ( \phi ) \triangleq 
		\Big[ 
		e^{\jmath\frac{2\pi}{\lambda}x_{1} \cos(\varphi_{\ell}^{(\rm r)} -\phi ) },  
		\ldots
		e^{\jmath\frac{2\pi}{\lambda}x_{N} \cos(\varphi_{\ell}^{(\rm r)} -\phi ) }
		\Big]^T, 
	\end{align}
	where $x_{n}$ denotes the horizontal coordinate of the $n$-th reflecting element along the $x$-axis with $n \in \mathcal{N}$, and $\varphi_{\ell}^{(\rm r)}$ is the angle-of-arrival (AoA) of the $\ell$-th path. Additionally, the corresponding PRM is given by 
	\begin{align}
		\mathbf{\Sigma} =\diag (\beta_{0},\beta_{1},\ldots,\beta_{L} ), 
	\end{align}
	where $\beta_{0} $ is the path-response coefficient of the LoS path and $\beta_{\ell} \sim \mathcal{CN}(0, \sigma_{\ell}^2),\ell \in \{1,2,\ldots, L\}$ is the complex-valued path-response coefficient of the $\ell$-th NLoS path.
	
	Similar to the BS-IRS channel $	\mathbf{G}( \mathbf{q},\boldsymbol{\zeta}) $,  the channel from the IRS to the $k$-th user can be modeled as
	\begin{align}
		\mathbf{r}_{k}(\phi) &= \sum_{\ell_{{\rm r},k}=0}^{L_{{\rm r},k}} \bar{\beta}_{k,\ell_{{\rm r},k}} \bar{\mathbf{a}}_{k,\ell_{{\rm r},k}} (\phi) 
		= \mathbf{A}_{{\rm r},k}(\phi )\boldsymbol{\beta}_{{\rm r},k}.
	\end{align}
	Specifically, the FRM and path-response vector (PRV) are given by
	\begin{align}
		\mathbf{A}_{{\rm r},k}(\phi) & = \big[\bar{\mathbf{a}}_{k,0} (\phi ),\ldots,\bar{\mathbf{a}}_{k,{L_{{\rm r},k}}} (\phi) \big] \in \mathbb{C}^{N \times (L_{{\rm r},k }+1) }, \\
		\boldsymbol{\beta}_{{\rm r},k} &= 
		\big[\bar{\beta}_{{k},0},\bar{\beta}_{{k},1},\ldots,\bar{\beta}_{{k},L_{{\rm r},k}} \big]^T \in \mathbb{C}^{L_{{\rm r},k}+1}, 
	\end{align}
	respectively, where the corresponding FRV of the $\ell_{{\rm r},k}$-th path is given by
	\begin{align}
		\bar{\mathbf{a}}_{{k},{\ell_{{\rm r},k}}} (\phi) & = \Big[e^{-\jmath\frac{2\pi}{\lambda}x_{1}\cos(\varphi_{k,{\ell_{{\rm r},k}}}^{(\rm r)} +\phi)  },\ldots, \nonumber \\
		&\quad \quad \quad e^{-\jmath\frac{2\pi}{\lambda}x_{N}\cos(\varphi_{k,{\ell_{{\rm r},k}}}^{(\rm r)} +\phi)  }
		\Big]^T \in \mathbb{C}^{N \times 1 },
	\end{align}
	with $\varphi_{k,{\ell_{{\rm r},k}}}^{(\rm r)}$ denoting the corresponding AoD and $L_{{\rm r},k}+1$ being the total number of paths between the IRS and user $k$.
	Additionally,  $\bar{\beta}_{{k},0}$ represents the path-response coefficient of the LoS path from the IRS to user $k$ and $\bar{\beta}_{{k},\ell_{{\rm r},k}} \sim \mathcal{CN}(0, \bar{\sigma}_{k,\ell_{{\rm r},k}}^{2} ),\ell_{{\rm r},k} \in \{1,2,\ldots,L_{{\rm r},k}\} $ represents the path-response coefficient of NLoS path from the IRS to user $k$.
	
	For the BS-user $k$ channel, we let $ L_{{\rm t},k}+1 $ denote the total number of paths   and  $\varphi_{k,{\ell_{{\rm t},k}}}^{(\rm t)}$ denote the corresponding AoD of the  $\ell_{{\rm t},k}$-th path. Then, the direct channel from the BS to the $k$-th user can be similarly modeled as
	\begin{align}
		\mathbf{h}_{k}(\mathbf{q},\psi)  \!=\!\!\!  \sum_{\ell_{{\rm t},k}=0}^{L_{{\rm t},k}} \tilde{\beta}_{k,\ell_{{\rm t},k}} \tilde{\mathbf{a}}_{k,\ell_{{\rm t},k}}(\mathbf{q},\psi ) 
		\!=\! \mathbf{A}_{{\rm t},k}  ( \mathbf{q},\psi )\boldsymbol{\beta}_{{\rm t},k},
	\end{align}
	with $\mathbf{A}_{{\rm t},k} ( \mathbf{q},\psi) $, $ \boldsymbol{\beta}_{{\rm t},k} $, and $  \tilde{\mathbf{a}}_{k,{\ell_{{\rm t},k}}}(\mathbf{q},\psi) $  denoting its
	corresponding FRM, PRV, and FRV, respectively.  Additionally,  $\tilde{\beta}_{k,0}$ represents the path-response coefficient of LoS path, and $\tilde{\beta}_{k,\ell_{{\rm t},k}} \sim \mathcal{CN}(0, \tilde{\sigma}_{k,\ell_{{\rm t},k}}^{2}) $ represents the  path-response coefficient of the $\ell_{{\rm t},k}$-th NLoS path.
	\vspace{-10pt}
	\subsection{Transmission Model} 
	The BS sends $K$ independent data streams to the $K$ users simultaneously with $s_{k} \sim \mathcal{CN}(0,1)$ denoting the transmitted signal to the $k$-th user. Then,  the received signals at the $k$-th user can be expressed as
	\begin{align}
		&y_{k} = \big( \mathbf{h}^H_{k}(\mathbf{q},\psi) + \mathbf{r}^H_{k} (\phi)  \mathbf{\Theta} \mathbf{G}( \mathbf{q},\boldsymbol{\zeta}) \big)  \sum_{i\in\mathcal{K}} \mathbf{w}_{i}  s_{i} +z_{k} \nonumber \\
		&=\big( \mathbf{h}^H_{k}(\mathbf{q}, \psi ) \!+\! \mathbf{g}^H_{k}(\mathbf{q},\boldsymbol{\zeta},\mathbf{\Theta}) \big) \sum_{i\in\mathcal{K}} \mathbf{w}_{i} s_{i}\!+\! z_{k},\forall k\in \mathcal{K},
	\end{align}
	where $ \mathbf{g}^H_{k}(\mathbf{q},\boldsymbol{\zeta},\mathbf{\Theta} ) \triangleq  \mathbf{r}^H_{k} (\phi)  \mathbf{\Theta} \mathbf{G}( \mathbf{q},\boldsymbol{\zeta}) $,
	$\mathbf{w}_{k}$ denotes the transmit beamforming vector for the $k$-th user, $\mathbf{\Theta} \triangleq \diag(\mathbf{v}) $ is the reflection coefficient matrix of the IRS with $\mathbf{v} \triangleq [e^{\jmath\theta_{1}},e^{\jmath\theta_{2}},\ldots,e^{\jmath\theta_{N}}]^T$ denoting the reflection coefficient vector and $\theta_{n} \in (0,2\pi]$. Additionally, $z_{k}~\sim \mathcal{CN} (0,\sigma^{2}) $  is the received additive white Gaussian noise (AWGN) at user $k$ with $\sigma^{2}$ representing the noise power for all the $K$ users. As such, the achievable sum-rate of the $K$ users in bits/second/Hertz (bps/Hz) is given by
	\begin{align}\label{sum-rate}
		R_{\rm sum} = \sum_{k\in\mathcal{K}} \log_2 (1+\gamma_{k} ), 
	\end{align}
	where $ \gamma_{k} = \frac{ \left|  \left(  \mathbf{h}^H_{k}(\mathbf{q},\psi)+\mathbf{g}^H_{k}(\mathbf{q},\boldsymbol{\zeta},\mathbf{\Theta})  \right)   \mathbf{w}_{k} \right| ^2}{\sum_{i\in\mathcal{K} \setminus \{k\} } \left|  \left( \mathbf{h}^H_{k}(\mathbf{q},\psi )+\mathbf{g}^H_{k}(\mathbf{q},\boldsymbol{\zeta},\mathbf{\Theta} ) \right) \mathbf{w}_{i}   \right| ^2  + \sigma^{2} } $ represents the signal-to-interference-plus-noise ratio (SINR) at user $k$.
	
	\subsection{Transmission Protocol and Problem Formulation}
	\begin{figure}[t]
		\centering
		\includegraphics[width=0.4\textwidth]{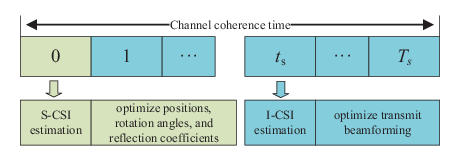}
		\caption{Transmission frame structure of the considered TTS transmission protocol.} 
		\label{Fig:Transframe}
		\vspace{-16pt}
	\end{figure}
	\subsubsection{Transmission Protocol}
	We consider a TTS transmission protocol~\cite{Zhaomm2021TSS,zhao2022secrecy}, as shown in Fig.~\ref{Fig:Transframe}. Specifically, in the long-timescale, S-CSI of all channels (including AoAs, AoDs, and the statistical distributions of path-response coefficients) is acquired by using existing channel estimation methods (see, e.g.,~\cite{MACE,XiaoMACE}). Based on the obtained S-CSI, the antenna configuration of 6DMA (including rotation angle $\psi$ and positions $\mathbf{q}$), as well as the rotation angle $\phi$ and reflection coefficient matrix $\mathbf{\Theta}$ of the R-IRS, are designed and set fixed during the channel coherence time, over which the S-CSI changes slowly or even remains unchanged~\cite{Adhikary_mmWave}. Next, in each short-timescale, the I-CSI of all channels, which refers to path-response coefficients, is obtained by using, e.g., the least-squares (LS) method~\cite{IRSCE2022}. As such, the transmit beamforming vectors are optimized to maximize the achievable sum-rate.
	
	This TTS beamforming design enjoys several advantages.
	First, it reduces the substantial channel estimation overhead associated with real-time I-CSI estimation. Second, compared to schemes that continuously adjust antenna and surface configurations, which are often ineffective, the TTS transmission significantly lowers implementation complexity. Last, the TTS design reduces the computational complexity involved in optimizing ${{\mathbf{q}, \boldsymbol{\zeta}, \mathbf{\Theta}, \mathbf{w}_k}}$ at each short-timescale, making it a more efficient and practical solution for the proposed system.

	\subsubsection{Problem Formulation}Under the above TTS transmission protocol, we aim at maximizing the average sum-rate of all users by jointly optimizing the transmit beamforming matrix in each short-timescale based on I-CSI (i.e., $\mathbf{H}_{k}(\mathbf{q},\boldsymbol{\zeta} )  \triangleq\{\mathbf{h}_{k}(\mathbf{q},\psi),\mathbf{r}_{k}(\phi),\mathbf{G}(\mathbf{q},\boldsymbol{\zeta}) \}$), as well as antenna/surface configurations (including 6DMA position, rotation and R-IRS rotation) and IRS reflection coefficients in the long-timescale based on S-CSI. Let $d = \frac{\lambda}{2}$ denote the minimum spacing between adjacent antennas and $\mathcal{C}_{\rm q} = [q_{\min}, q_{\max}]$ represent the allowable movement region with $D = q_{\max}-q_{\min}$ denoting its maximal aperture. 
	Additionally, $ \mathcal{C}_{\psi} = [\psi_{\min},\psi_{\max}]$ and $ \mathcal{C}_{\phi} = [\phi_{\min},\phi_{\max}]$ represent the allowable rotation angle regions of R-IRS and 6DMA, where $\psi_{\min}$,  $\psi_{\max}$,  $\phi_{\min}$, and $\phi_{\max}$ denote their minimum and maximum rotation angles, respectively.
	Based on the above, this optimization problem can be formulated as
	\begin{subequations}
		\begin{align}
			(\textbf{P1}):\; \max_{\mathbf{q},\boldsymbol{\zeta},\mathbf{\Theta}}&\quad\mathbb{E} \big[ \max_{\{\mathbf{w}_{k}\}}~\sum_{k\in\mathcal{K}} \log_2 (1+\gamma_{k})   \big] \nonumber \\
			{\rm {s.t.}}&\quad\theta_{n} \in (0,2\pi],~n\in\mathcal{N},\label{C_PS} \\
			&\quad|q_{i}-q_{j}| \ge d,~1\le i\neq j \le M,\label{C_MA1} \\
			&\quad q_{m} \in \mathcal{C}_{\rm q},~m \in \mathcal{M},\label{C_MA2}  \\
			&\quad \psi \in \mathcal{C}_{{\psi}},\label{M_Rot}\\
			&\quad \phi \in \mathcal{C}_{{\phi}},\label{C_Rot} \\
			&\quad  \sum_{k\in\mathcal{K}} ||\mathbf{w}_{k}||_2^{2} \le P_{\rm t}\label{C_Pt}.
		\end{align} 
	\end{subequations}
	Specifically, constraint~\eqref{C_PS} specifies the phase shifts of R-IRS, 
	constraints~\eqref{C_MA1} and~\eqref{C_MA2} enforce the minimum antenna spacing and allowable movement region,
	constraints~\eqref{M_Rot} and~\eqref{C_Rot} regulate the allowable rotation regions of 6DMA and R-IRS,
	and constraint~\eqref{C_Pt} restricts the maximum transmit power. 
	
	Problem (\textbf{P1}) is a non-convex optimization problem, which is generally difficult to solve optimally due to the following reasons: 1) the coupling of multiple optimization variables in the objective function; 2) the non-convex expressions for the rotation angle and positions of 6DMA, as well as the rotation angle of R-IRS; 3) and the lack of a closed form expression for the average sum-rate.
	In the following section, we first consider a single-user scenario to obtain useful insights, for which an efficient algorithm is proposed to solve this non-convex problem. Then we further extend the results to the general multi-user scenario.
	
	\section{Single-user Scenario}\label{Sec3}
	In this section, we focus on the single-user scenario (i.e., $K=1$), and solve the average rate maximization problem based on the TTS transmission protocol. 
	Accordingly, the optimization Problem (\textbf{P1}) can be simplified as 
	\begin{subequations}
		\begin{align}
			(\textbf{P2}):\; \max_{\mathbf{q},\boldsymbol{\zeta},\mathbf{\Theta}}&\quad\mathbb{E} \big[ \max_{\mathbf{w}_{1}}~ \log_2 (1+\gamma_{1})   \big] \nonumber  \\
			{\rm {s.t.}}&\quad\eqref{C_PS}-\eqref{C_Rot}, \nonumber\\
			&\quad \|\mathbf{w}_{1}\|_2^{2} \le P_{\rm t}.\label{C_Pt_single}
		\end{align} 
	\end{subequations}
	For the short-timescale optimization problem, the optimal transmit beamforming in each I-CSI is obtained in closed form. Then, for the long-timescale optimization problem, the optimization variables in the objective function are decoupled based on S-CSI, and an efficient algorithm is proposed to solve this non-convex optimization problem. Finally, some special cases are considered, based on which we analytically demonstrate the effectiveness of R-IRS and 6DMA in improving rate performance.  
	
	\subsection{Proposed Solution to Single-user Scenario}
	
	\subsubsection{Short-timescale Optimization}
	In each short-timescale, the transmit beamforming at the BS is optimized based on I-CSI, i.e., $ \{\mathbf{h}_{1}(\mathbf{q},\psi),\mathbf{r}_{1}(\phi),\mathbf{G}(\mathbf{q},\boldsymbol{\zeta}) \} $. Similar to~\cite{Zhaomm2021TSS}, the optimal solution to the short-timescale optimization Problem (\textbf{P2}) is maximum-ratio transmission (MRT), which is given by 
	\begin{align}\label{closed_tbv_SU}
		\mathbf{w}_{1} = \sqrt{P_{\rm t}} \frac{ \mathbf{h}_{1}(\mathbf{q},\psi)+\mathbf{g}_{1}(\mathbf{q},\boldsymbol{\zeta},\mathbf{\Theta})  }{\|   \mathbf{h}_{1}(\mathbf{q},\psi)+\mathbf{g}_{1}(\mathbf{q},\boldsymbol{\zeta},\mathbf{\Theta}) \|_2}.
	\end{align}

	\subsubsection{Long-timescale Optimization}
	Based on the optimized transmit beamforming obtained in~\eqref{closed_tbv_SU},
	we then optimize the long-timescale related variables $\{\mathbf{q},\boldsymbol{\zeta},\mathbf{\Theta} \}$ based on S-CSI. 
	By leveraging Jensen's inequality, which provides an approximation of the average rate~\cite{Wang_Jensen}, the long-timescale optimization problem can be formulated as
	\begin{align}
		(\textbf{P3}):\; \max_{\mathbf{q},\boldsymbol{\zeta},\mathbf{\Theta}}&\quad\mathbb{E} \Big[  \big\| \mathbf{h}_{1}(\mathbf{q},\psi )+\mathbf{g}_{1}(\mathbf{q},\boldsymbol{\zeta},\mathbf{\Theta}) \big\|_2^2   \Big] \nonumber  \\
		{\rm {s.t.}}&\quad\eqref{C_PS}-\eqref{C_Rot}, \nonumber
	\end{align} 
	where the expectation is taken over I-CSI, i.e., $\mathbf{H}_{1}(\mathbf{q},\boldsymbol{\zeta} )$.
	The objective function of Problem (\textbf{P3}) is non-convex due to the quadratic expression under the expectation operation, which makes it challenging to solve directly. To address this issue, we first rewrite the objective function of Problem (\textbf{P3}) as follows based on S-CSI.
	\begin{lemma}[Expected equivalent channel power gain]\label{closed_form1}
		\rm 
		Based on S-CSI,	the objective function of Problem (\textbf{P3}) can be equivalently rewritten~as 
		\begin{subequations}
			\begin{align}
				&~~~~\mathbb{E}\Big[  \big\| \mathbf{h}_{1}(\mathbf{q},\psi )+\mathbf{g}_{1}(\mathbf{q},\boldsymbol{\zeta},\mathbf{\Theta}) \big\|_2^2   \Big]  \nonumber \\
				&=\mathbb{E} \big[ \mathbf{h}_{1}^H(\mathbf{q},\psi) \mathbf{h}_{1}(\mathbf{q},\psi)\big] 
				+\mathbb{E} \big[ \mathbf{g}_{1}^H(\mathbf{q},\boldsymbol{\zeta},\mathbf{\Theta} ) \mathbf{g}_{1}(\mathbf{q},\boldsymbol{\zeta},\mathbf{\Theta}) \big] \nonumber \\
				&~~~+ \mathbb{E} \big[2\mathcal{R} \big\{ \mathbf{h}_{1}^H(\mathbf{q},\psi)  \mathbf{g}_{1}(\mathbf{q},\boldsymbol{\zeta},\mathbf{\Theta}) \big\}\big]   \label{derivation}\\
				&=\!\underbrace{c_1}_{\text{1st}} \!\!+ \underbrace{\mathbf{v}^T \hat{\mathbf{G}}_{1} (\phi)  \mathbf{v}^\dagger}_{\text{2nd}} \!+\! \underbrace{2\mathcal{R} \big\{ \omega_{1} \mathbf{v}^T  \hat{\mathbf{a}}_{{1},{0}} (\phi )
					\hat{a}_{{1},{0}}(\mathbf{q},\psi )\big\}}_{\text{3rd}}. \label{obj_closed1}
			\end{align}
		\end{subequations}
	\end{lemma}
	\begin{proof}
		Please refer to Appendix~\ref{App1}.
	\end{proof}
	
	{\bf Lemma~\ref{closed_form1}} provides a closed-form expression for the objective function of Problem (\textbf{P3}), thereby making it possible to directly solve Problem (\textbf{P3}). 
	Moreover, one can observe that the first term of~\eqref{obj_closed1} (i.e., $c_{1}$) is the expected direct path power gain, which is constant; the second term is the expected reflect path power gain, which is related to the IRS rotation ($\phi$) only; while the third term is the expectation of the correlation between direct and reflect paths, which is associated with both IRS rotation ($\phi$), 6DMA rotation ($\psi$), and 6DMA position ($\mathbf{q}$). The coupling of multiple optimization variables in the third term makes it challenging to obtain the optimal solution. Fortunately, the following {\bf Lemma~\ref{decoupled}} can be utilized to decouple the optimization variables. 
	\begin{lemma}\label{decoupled}
		\rm
		The expected equivalent channel power gain in~\eqref{obj_closed1} can be rewritten as
		\begin{align}
			c_1 \!+\! \mathbf{v}^T \hat{\mathbf{G}}_{1}(\phi)  \mathbf{v}^\dagger + 2 |\omega_{1}|  |\hat{a}_{{1},{0}}(\mathbf{q},\psi)|  \mathcal{R}  \big\{  \mathbf{v}^T   \hat{\mathbf{a}}_{{1},{0}} (\phi)
			\big\}. \label{obj_closed2}
		\end{align} 
	\end{lemma}
	\begin{proof}
		Please refer to Appendix~\ref{Decoupl_app}.
	\end{proof}
	
	Based on {\bf Lemma~\ref{decoupled}}, it is apparent that the optimizations of the rotation angle and positions of 6DMA are decoupled from the optimizations of the rotation angle and reflection coefficient vector of IRS. As a result, we can decompose the long-timescale Problem (\textbf{P3}) into two parallel subproblems, corresponding to subproblem 1 for optimizing the rotation angle and positions of 6DMA \{$\psi,\mathbf{q}$\}, and subproblem 2 for optimizing the rotation angle and reflection coefficient vector of IRS \{$\phi, \mathbf{v}$\}. In the following, the solutions to these two subproblems are presented, respectively.
	
	\textbf{\underline{Subproblem 1}:}
	Since the rotation angle and reflection coefficients of R-IRS have no effect on optimizing the rotation angle and positions of 6DMA, the subproblem of optimizing 6DMA positions and rotation can be simplified as
	\begin{align}
		(\textbf{P3.1}):\; \max_{\mathbf{q},\psi}\quad& |\hat{a}_{{1},{0}}(\mathbf{q},\psi)|  \nonumber  \\
		{\rm {s.t.}}\quad&\eqref{C_MA1},\eqref{C_MA2}. \nonumber
	\end{align} 
	Due to the constraints on allowable movement region (i.e.,~\eqref{C_MA2}) and rotation angle (i.e.,~\eqref{M_Rot}), it is hard to directly obtain the optimal solution to Problem (\textbf{P3.1}). To tackle the coupling among the variables $\{ \mathbf{q},\psi \}$ in the objective function of Problem (\textbf{P3.1}), we employ the exhaustive search method to find the optimal 6DMA rotation. For each 6DMA rotation angle $\psi \in [\psi_{\min},\psi_{\max}]$, we then consider two different cases to investigate the impact of the allowable movement region on system performance. 
	\begin{case}[Infinite movement region]
		\rm
		In the case of an infinite movement region, we can remove constraint~\eqref{C_MA2} in Problem (\textbf{P3.1}). Similar to~\cite{Zhou_MA},  given the rotation angle of 6DMA~$\psi$, after removing constraint~\eqref{C_MA2}, the optimal solution to Problem (\textbf{P3.1}) can be obtained as follows.
		\begin{lemma}[Optimal positions of MAs]\label{Lem:optPOS}
			\rm  The optimal solution to Problem (\textbf{P3.1}) without the movement region constraint \eqref{C_MA2} is given by $\mathbf{q}_1 = [q_1,\ldots,q_{M}]^T$, with the $m$-th element of which satisfying
			\begin{align}\label{closed_MA}
				q_m = q_{\min} +\frac{m-1}{\Delta(\varphi_{0}^{(\rm t)}, \varphi_{1,0}^{(\rm t)},\psi)}\lambda,~m \in \mathcal{M},
			\end{align}
			where $\Delta(\varphi_{0}^{(\rm t)}, \varphi_{1,0}^{(\rm t)},\psi) = \big|\cos(\varphi_{0}^{(\rm t)} + \psi)  + \cos(\varphi_{1,0}^{(\rm t)} - \psi)\big|$.
		\end{lemma}	
		\begin{proof}
			This lemma can be proved by using a similar method as in [Theorem 1,~\cite{Zhou_MA}].
		\end{proof}
	
		Note that the MA positions obtained in {\bf Lemma~\ref{Lem:optPOS}} realize a \emph{uniform sparse array}, for which the BS generates two separate beams by exploiting sparse-array grating lobs, hence, enabling favorable phase alignment between the direct and reflected links. 
		With the solution given in~\eqref{closed_MA},  $|\hat{a}_{{1},{0}}(\mathbf{q},\psi)|$ can be upper-bounded as $|\hat{a}_{{1},{0}}(\mathbf{q},\psi)| \le M$. Under this condition, the aperture of 6DMA array is $  D_{1} = {(M-1)\lambda}/{\Delta(\varphi_{0}^{(\rm t)}, \varphi_{1,0}^{(\rm t)},\psi)} $. In contrast, for FPA system, achieving coordination between the direct and reflected links comes at the cost of array gain, resulting in inferior performance.
	\end{case}

	\begin{case}[Finite movement region]  
		\rm 
		Although the optimal position in~\eqref{closed_MA} ensure the maximal correlation between $\mathbf{a}_{{\rm t},0}( \mathbf{q},\psi)$ and $ \tilde{\mathbf{a}}_{1,{0}} (\mathbf{q},\psi )$, it may not meet constraint~\eqref{C_MA2} (i.e., $ D_{1} > D$) when $\Delta(\varphi_{0}^{(\rm t)}, \varphi_{1,0}^{(\rm t)},\psi)$ is relatively small. In other words, when $\Delta(\varphi_{0}^{(\rm t)}, \varphi_{1,0}^{(\rm t)},\psi) < \frac{(M-1)\lambda}{D}$, the solution given in~\eqref{closed_MA} no longer holds due to the finite movement region constraint. 
		
		\begin{remark}[Conventional optimization based method]
			\rm 
			In fact, the optimization-based method can be utilized to optimize the positions of 6DMA by using, e.g., Taylor expansion to handle the non-convex expressions in both the objective function and constraints. However, it introduces the following issues. First, the objective function of Problem (\textbf{P3.1}) is determined by multiple coupled variables, resulting in considerable computational complexity. Second, the optimization-based method is prone to falling into a local optimum or even a low-quality solution, thus degrading the performance~\cite{ZhuMAtutorial}.
			\vspace{-6pt}
		\end{remark}
		Therefore, to solve Problem (\textbf{P3.1}) with $D_{1} > D$, we propose an efficient DE algorithm~\cite{MAcovert} to obtain a suboptimal solution to Problem (\textbf{P3.1}). 
		Specifically, in the proposed DE algorithm, the positions of 6DMA are updated by the following three operations: mutation, crossover, and selection. The operations of mutation and crossover are intended to generate trial positions for 6DMA, and the selection is devised to evaluate the quality of these trial positions through the fitness function. The details of the proposed DE algorithm are presented as follows. 
		
		At the beginning of DE algorithm, a population of $P$ individuals is randomly generated, which are expressed as 
		\begin{align}
			\mathcal{P}^{(0)}= \{\mathbf{q}_{1}^{(0)},\mathbf{q}_{2}^{(0)},\ldots,\mathbf{q}_{P}^{(0)}\},~p\in\{1,2,\ldots,P\},
		\end{align}
		where $\mathbf{q}_{p}^{(0)}$ denotes the initial position of the $p$-th individual. Note that the quality of each individual is evaluated by the following fitness function
		\begin{align}\label{fitness_SU}
			\mathcal{F}_{1} ( \mathbf{q}_{p}^{(s)} ) = \big|\hat{a}_{1,{0}} (\mathbf{q}_{p}^{(s)} ) \big| - \eta \mathcal{B}_1(\mathbf{q}_{p}^{(s)} )  \big| \mathcal{B}_2(\mathbf{q}_{p}^{(s)} ) \big|,
		\end{align}
		where $\mathbf{q}_{p}^{(s)} $ represents the $p$-th individual in the $s$-th iteration. The first term of~\eqref{fitness_SU} is the objective function of Problem (\textbf{P3.1}), while the second term represents the penalty introduced to ensure the minimum antenna spacing constraint, with $ \eta $ serving as a scaling factor. Additionally,  $\mathcal{B}_1(\mathbf{q})$ represents the situation of constraint violation, which is defined as
		\begin{align}
			\mathcal{B}_1(\mathbf{q}) = \sum_{\mathbf{q}\in\mathcal{B}_2(\mathbf{q} )} ( d - |q_i-q_j| ), 
		\end{align}
		$\mathcal{B}_2(\mathbf{q})$ refers to the set whose elements do not adhere to constraint~\eqref{C_MA1}, as defined by
		\begin{align}
			\mathcal{B}_2(\mathbf{q} ) = \Big\{ (q_{i},q_{j})\Big|~ | q_{i}-q_{j} |< d, 1\le i<j\le M   \Big\}.
		\end{align}
		Based on the generated individual (i.e., positions of 6DMA) in each iteration, the corresponding value of the fitness function and the optimal individual in the current population $(\textrm {i.e.},  \mathbf{q}_{\rm best}^{(s)} = \mathop{\arg\max}_{(p)} \mathcal{F}_{1} ( \mathbf{q}_{p}^{(s)} ) )$ can be obtained. Then, the process of individual evolution, involving three operations, is performed to update the positions of 6DMA, which are respectively given as follows.
		
		\textbf{Mutation:} The mutation operation is conducted to produce a mutant individual with some differential individual, which is given by\footnote{The DE/best/1 mutation operation is employed here, showcasing its potential to achieve sub-optimal performance in the single-user scenario.}
		\begin{align}\label{mutation}
			\mathbf{m}_{p}^{(s)} = \mathbf{q}_{\rm best}^{(s-1)} + F (\mathbf{q}_{r_1}^{(s-1)} -\mathbf{q}_{r_2}^{(s-1)} ),
		\end{align}
		where $\mathbf{q}_{r_1}^{(s-1)}$ and $\mathbf{q}_{r_2}^{(s-1)}$ are two different individuals randomly chosen from the current population (i.e., $ \mathcal{P}^{(s-1)}$) with $r_1$ and $r_2$ satisfying $r_1 \neq r_2 \neq p$. Additionally, the mutation factor $F$ can be adjusted to control the global search capability and the convergence rate of DE algorithm.
		
		\textbf{Crossover:} After mutation operation, the crossover operation is employed to facilitate genetic exchanges between the mutant individual and current individual for generating a trial individual. The process of crossover operation can be mathematically expressed as
		\begin{equation}\label{crossover}
			\mathbf{u}_{p}^{(s)} = \left\{
			\begin{aligned}
				&\mathbf{m}_{p}^{(s)},~&&\textrm{if}~ {\rm rand}(1) < C_{\rm R}~\textrm{or}~p =P_{\rm rand}^{(s)}, \\
				&\mathbf{q}_{p}^{(s-1)},~&& \textrm{otherwise},
			\end{aligned}
			\right.
		\end{equation}
		where ${\rm rand}(1)$ represents a random variable following a uniform distribution between $0$ and $1$, and $P_{\rm rand}^{(s)}$ denotes a random positive integer not exceeding $P$ in the $s$-th iteration to ensure the occurrence of at least one crossover operation. Additionally, $C_{\rm R}$ is the crossover factor for controlling population diversity. 
		Furthermore, the following operation is executed to satisfy constraint~\eqref{C_MA2}.
		\begin{align}\label{DE_SU_EE}
			\big[\mathbf{u}_{p}^{(s)}\big]_{m} = \max\big\{ \min\big\{\big[\mathbf{u}_{p}^{(s)}\big]_{m}, q_{\rm max}\big\}, q_{\rm min} \big\}.
		\end{align}
		
		\textbf{Selection:} After crossover operation, the trial individual is evaluated using the fitness function to generate the next population, as determined by
		\begin{equation}\label{select} 
			\mathbf{q}_{p}^{(s)}=\left\{
			\begin{aligned}
				&\mathbf{u}_{p}^{(s)}, &&\textrm{if}~~ \mathcal{F}_{1} ( \mathbf{u}_{p}^{(s)})> \mathcal{F}_{1} ( \mathbf{q}_{p}^{(s-1)} ),  \\
				&\mathbf{q}_{{p}}^{(s-1)}, &&\textrm{otherwise}.  \\
			\end{aligned}
			\right.
		\end{equation}
		After $S$ iterations, the suboptimal solution to Problem (\textbf{P3.1}), denoted as $\mathbf{q}_2$, can be obtained, satisfying $\mathbf{q}_2 = \mathbf{q}_{\rm best}^{(S)}$.
		\vspace{-4pt}
	\end{case}
	
	Based on the above, given the rotation angle of 6DMA, the solution to Problem (\textbf{P3.1}) is obtained as follows.
	\begin{proposition}[Proposed solution to Problem (\textbf{P3.1})]
		\rm Given rotation angle of 6DMA, the solution to Problem (\textbf{P3.1}) is given by
		\begin{equation}\label{Solution_t}
			{\mathbf{q}} =\left\{
			\begin{aligned}
				&\mathbf{q}_1, &&\textrm{if}~~\Delta(\varphi_{0}^{(\rm t)}, \varphi_{1,0}^{(\rm t)},\psi) > \frac{(M-1)\lambda}{q_{\max}-q_{\min}},  \\
				&\mathbf{q}_2, &&\textrm{otherwise}.
			\end{aligned}
			\right.
		\end{equation}
	\end{proposition}
	The exhaustive search method is employed to traverse all potential rotation angles of 6DMA and identify the maximum objective value achieved by solving (\textbf{P3.1}), thereby determining the optimal rotation angle and its corresponding positions of 6DMA $\{\psi^{*},\mathbf{q}^{*}\}$.
	
	\textbf{\underline{Subproblem 2}:}
	After obtaining the optimized rotation angle and positions of 6DMA, we then optimize the rotation angle and reflection coefficients of R-IRS. To address the coupling of $\{\phi, \mathbf{v}\}$ in~\eqref{obj_closed2}, we also employ the exhaustive search method to find the optimal IRS rotation angle, while for each rotation angle, the SDR technique is utilized to optimize~$\mathbf{v}$.
	Specifically, given the rotation angle of R-IRS $\phi \in \left[\phi_{\min},\phi_{\max} \right] $, the problem of optimizing IRS reflection coefficients can be formulated as  
	\begin{subequations}
		\begin{align}
			(\textbf{P3.2}):\; \max_{\mathbf{V}}&\quad \text{Tr} (   \mathbf{H}_{{\rm eff},1} \mathbf{V})  \nonumber  \\
			{\rm {s.t.}}&\quad \mathbf{V}(n,n) \le 1,~n\in\{1,2,,\ldots,N \},\label{V_1} \\
			&\quad \mathbf{V} (N+1,N+1) = 1, \label{V_2}\\
			&\quad \text{Rank}( \mathbf{V}) =1,\label{V_3}
		\end{align} 
	\end{subequations}
	where  $\mathbf{V} =\small \begin{bmatrix}
		\mathbf{v}^\dagger\\1   
	\end{bmatrix}  
	\begin{bmatrix}
		\mathbf{v}^T,  1
	\end{bmatrix}  $  and  $\mathbf{H}_{{\rm eff},1} = \small
	\begin{bmatrix}
		\hat{\mathbf{G}}_{1}(\phi)  & \mathbf{b}(\phi)  \\
		\mathbf{b}(\phi)^H           & c_1 
	\end{bmatrix} $ 
	with $ \mathbf{b} (\phi ) = \omega_{1}  \hat{a}_{{1},{0}}(\mathbf{q}^{*}, \psi^{*}) \hat{\mathbf{a}}_{{1},{0}} (\phi) $.
	Since Problem (\textbf{P3.2}) is a semidefinite program (SDP) problem, it can be solved by using e.g., CVX tool~\cite{grant2014cvx} after relaxing the rank-one constraint in~\eqref{V_3}. After traversing all potential rotation angles of R-IRS, we thus identify the maximum objective value by solving Problem (\textbf{P3.2}). Based on the above, the optimal rotation angle and  reflection coefficients of R-IRS are obtained. 
	\vspace{-12pt}
	\subsection{Discussions and Analysis}\label{Sec3-B}
	\subsubsection{Discussions} In this subsection, we focus on the single-user and single-path scenario to draw some useful insights into the proposed joint R-IRS and 6DMA-BS architecture. When $L = 0 $,~\eqref{obj_closed2} in {\bf Lemma~\ref{decoupled}} can be simplified as 
	\begin{align}
		&~c_1 + \mathbf{v}^T \hat{\mathbf{G}}_{1}(\phi)  \mathbf{v}^\dagger + 2 |\omega_{1}|  |\hat{a}_{{1},{0}}(\mathbf{q},\psi)|  \mathcal{R}  \big\{  \mathbf{v}^T   \hat{\mathbf{a}}_{{1},{0}} (\phi)
		\big\}   \nonumber \\
		= &~c_1 +  M |\beta_0|^2 |\bar{\beta}_{{1},0}|^2   \mathbf{v}^T \big( \hat{\mathbf{a}}_{{1},{0}} (\phi) \hat{\mathbf{a}}^H_{{1},{0}} (\phi)\big) \mathbf{v}^\dagger   \nonumber \\ 
		& +  2 |\omega_{1}|  |\hat{a}_{{1},{0}}(\mathbf{q},\psi)|  \mathcal{R}  \big\{  \mathbf{v}^T   \hat{\mathbf{a}}_{{1},{0}} (\phi) 
		\big\} \nonumber \\
		\overset{(g)}{=} &~c_1 +  M N^ 2|\beta_0|^2 |\bar{\beta}_{{1},0}|^2 +2 N |\omega_{1}|  |\hat{a}_{{1},{0}}(\mathbf{q},\psi)|, \label{closed_form3}
	\end{align}
	where $ (g) $ holds when $\mathbf{v}^T = \hat{\mathbf{a}}^H_{{1},{0}} (\phi)$. 
	It is observed that in this scenario, the expected equivalent channel power gain is independent of the R-IRS rotation angle. Consequently, in the single-path case, the rotation of R-IRS does not contribute to system performance improvement, as it provides no additional gain in the expected equivalent channel power in~\eqref{closed_form3}.
	
	\begin{remark}[When 6DMA-BS, R-IRS, or both are needed?]\label{Rem2}
		\rm  Based on {\bf Lemma~\ref{decoupled}}, it is evident that both the rotation angle and positions of 6DMA as well as the rotation angle of IRS have an impact on~\eqref{obj_closed2}, indicating that both 6DMA and R-IRS are beneficial in the generic multi-path scenario (i.e., $L= {L}_{{\rm r},k}={L}_{{\rm t},k} >0$ ). For the LoS-only scenario (i.e., $L = {L}_{{\rm r},k} = {L}_{{\rm t},k} = 0$), the R-IRS provides little improvement in user rate performance, whereas the 6DMA can still yield significant gains. On the other hand, in a multi-path scenario, the rotation of the R-IRS can enhance the expected equivalent reflected link power by adjusting the rotation angle for multi-path alignment, 
		thereby improving the average rate performance.
	\end{remark}
	\subsubsection{Algorithm Convergence}
	For the proposed algorithm, its convergence is mainly determined by the optimization of 6DMA positions. Considering the fact that in each DE iteration, we have
	$	\mathcal{F}_{1} ( \mathbf{q}_{\rm best}^{(s)} ) \ge \mathcal{F}_{1} ( \mathbf{q}_{\rm best}^{(s-1)} )$,
	and the upper bound of objective value in Problem (\textbf{P3.1}) is $M$. Thus, the convergence of DE algorithm can be guaranteed. In addition, the 1D exhaustive search based method for solving subproblem 2 ensures that the expected equivalent channel gain in~\eqref{obj_closed2} is non-decreasing. Based on the above, the convergence of proposed algorithm can be established.

	\subsubsection{Computational Complexity} 
	Note that the computational complexity of the proposed algorithm
	is determined by the optimizations of subproblems 1 and 2. Specifically, the computational complexity of solving subproblem 1 is in the order of $\mathcal{O}(PS)$, while that of solving subproblem 2 is in the order of $\mathcal{O}\big((N+1)^{4.5} \log(1/ \epsilon)\big)$~\cite{MyTWC}, where $\epsilon$ is the required accuracy of CVX solver. Thus, the overall complexity of proposed algorithm
	is in the order of  $\mathcal{O}(  T_{\phi}  (N+1)^{4.5} \log(1/ \epsilon) + T_{\psi} PS) $, where $T_{\phi}$ and $T_{\psi}$ denote the number of 1D exhaustive search for $\phi$ and $\psi$, respectively.

	\vspace{-6pt}
	\section{Multi-user Scenario}\label{Sec4:label}
	In this section, we aim to maximize the average sum-rate for the multi-user scenario. Specifically, in each short-timescale, we optimize the transmit beamforming of the BS based on I-CSI. Then, in the long-timescale, an extended DE algorithm is devised to optimize the rotation angle and positions of 6DMA, as well as the rotation angle and reflection coefficients of the IRS based on S-CSI.
	
	\subsection{Proposed Solution to Multi-user Scenario}\label{Sec:IV}
	
	\subsubsection{Short-timescale Optimization}\label{Sec:IV-A}
	In each short-timescale slot $t \in\{1,\ldots,T\}$, the BS designs its transmit beamforming vectors $\{\mathbf{w}_{k}\}$ based on I-CSI, denoted by $\mathbf{H}_{k}(\mathbf{q},\boldsymbol{\zeta},t)  \triangleq\{\mathbf{h}_{k}(\mathbf{q},t) ,\mathbf{r}_{k}(\phi,t) ,\mathbf{G}(\mathbf{q},\boldsymbol{\zeta},t) \}$. The corresponding optimization problem can be formulated as 
	\begin{align}
		(\textbf{P4}):\;  \max_{\mathbf{w}_{k}}&\quad \sum_{k\in\mathcal{K}} \log_2 \Big( 1+\gamma_{k}\big(\mathbf{H}_{k}(\mathbf{q},\boldsymbol{\zeta},t), \mathbf{\Theta}  \big)   \Big)   \nonumber \\
		{\rm {s.t.}}
		&\quad \eqref{C_Pt}. \nonumber
	\end{align} 
	Note that Problem (\textbf{P4}) is a weighted sum-rate maximization problem with the weight set as $1$. Thus, we design the transmit beamforming of the BS by iteratively updating $\{\mathbf{w}_{k}\}$  via using the WMMSE techniques~\cite{Guo2020WSR}
	\begin{subequations}\label{MMSE_closed}
		\begin{align}
			\chi_{k} &= \big(\sum_{i\in\mathcal{K}} | \mathbf{h}_{{\rm eff},k}^H \mathbf{w}_{i} |^2 +\sigma^2  \big)^{-1} \mathbf{h}_{{\rm eff},k}^H \mathbf{w}_{k},\label{MMSE1} \\
			\kappa_{k} & = (1- \chi_{k}^{\dagger}\mathbf{h}_{{\rm eff},k}^H \mathbf{w}_{k})^{-1}, \label{MMSE2}\\
			\mathbf{w}_{k} &= \chi_{k} \kappa_{k} \big(\mu \mathbf{I}_{M} + \sum_{i\in\mathcal{K}} |\chi_{i}|^2 \kappa_{i} \mathbf{h}_{{\rm eff},i}\mathbf{h}_{{\rm eff},i}^H \big)^{-1} \mathbf{h}_{{\rm eff},k},\label{MMSE3} 
		\end{align}
	\end{subequations}
	where $\mathbf{h}_{{\rm eff},k} \triangleq \mathbf{h}_{k}(\mathbf{q},t)+ \mathbf{g}_{k}(\mathbf{q},\boldsymbol{\zeta},\mathbf{\Theta},t) $, $\mu \ge 0$ represents the  dual variable for the transmit power constraint, which can be obtained via bisection search.\footnote{Let $\mathbf{h}_{{\rm eff},k}(\mathbf{q},\boldsymbol{\zeta},\mathbf{\Theta},t)$ denote the short-timescale effective channel. We drop the arguments in $\mathbf{h}_{{\rm eff},k}$ here for simplicity.}
	With the optimized beamforming matrix of the BS, i.e., $ \{\mathbf{w}_k^*\} $, we obtain the objective function value for each short-timescale, which is expressed as 
	$R_{k} \big(\mathbf{H}(\mathbf{q},\boldsymbol{\zeta},t), \{\mathbf{w}_{k}(\mathbf{H}(\mathbf{q},\boldsymbol{\zeta},t), \mathbf{\Theta})\big\},\mathbf{\Theta})$ with $\mathbf{H}(\mathbf{q},\boldsymbol{\zeta},t) \triangleq\{\mathbf{H}(\mathbf{q},\boldsymbol{\zeta},t)\}_{k=1}^{K}$.
	
	\subsubsection{Long-timescale Optimization}\label{Sec:IV-A-2}
	In the long-timescale, our objective is to optimize the rotation angle and positions of 6DMA, as well as the rotation angle and reflecting coefficients of the IRS based on S-CSI, the optimization problem of which is formulated as
	\begin{align}
		(\textbf{P5}): \max_{\mathbf{q},\boldsymbol{\zeta},\mathbf{\Theta}}&~\mathbb{E} \Big[\sum_{k\in\mathcal{K}}  R_{k} \big(\mathbf{H}(\mathbf{q},\boldsymbol{\zeta},t), \big\{\mathbf{w}_{k}(\mathbf{H}(\mathbf{q},\boldsymbol{\zeta},t), \mathbf{\Theta})\big\},\mathbf{\Theta})  \Big] \nonumber  \\
		{\rm {s.t.}}&~\eqref{C_PS},\eqref{C_MA1},\eqref{C_MA2},\eqref{M_Rot},\eqref{C_Rot}. \nonumber
	\end{align} 
	Unlike the single-user scenario for which a closed-form expression for the transmit beamforming vector can be obtained in~\eqref{closed_tbv_SU}, there generally lacks of an explicit closed-form expression for the optimized BS beamforming vectors $\{\mathbf{w}_{k}( \mathbf{H}(\mathbf{q},\boldsymbol{\zeta},t ), \mathbf{\Theta} )\}$, posing new challenge to solve the long-term optimization problem. Moreover, the coupling of optimization variables $\{\mathbf{q},\boldsymbol{\zeta},\mathbf{\Theta} \}$ further complicates this problem. 
	
	To solve this long-timescale optimization problem, we propose to first transform the long-timescale optimization problem into two-layer optimization problems, corresponding to an inner problem for optimizing R-IRS reflection coefficient given fixed antenna/surface configurations $\{\mathbf{q},\boldsymbol{\zeta}\}$ as well as an outer problem for optimizing positions $\mathbf{q}$ and rotations $\boldsymbol{\zeta}$. The two problems are formulated and solved as follows.

	\textbf{\underline{Inner problem}:} Given any feasible antenna/surface configurations (including 6DMA position, rotation, and R-IRS rotation), Problem (\textbf{P5}) reduces to the optimization of reflection coefficient matrix, which is formulated as
	\begin{subequations}
		\begin{align}
			(\textbf{P5.1}):\; \max_{\mathbf{\Theta}}&\quad \mathbb{E} \Big[\sum_{k\in\mathcal{K}} R_{k} \big(\mathbf{H}(\boldsymbol{\varsigma},t),  \big\{\mathbf{w}_{k}\big( \mathbf{H}(\boldsymbol{\varsigma},t), \mathbf{\Theta}\big)\big\}, \mathbf{\Theta} \big)\Big]  \nonumber  \\
			{\rm {s.t.}}&\quad \eqref{C_PS}, \nonumber
		\end{align} 
	\end{subequations}
	where $\boldsymbol{\varsigma} \triangleq [\mathbf{q}^T,\boldsymbol{\zeta}^T]^T \in \mathbb{R}^{M+2}$. 
	Note that there generally lacks a closed-form expression for the average sum-rate, which is associated with $ R_{k} \big(\mathbf{H}(\boldsymbol{\varsigma},t),  \big\{\mathbf{w}_{k}\big( \mathbf{H}(\boldsymbol{\varsigma},t), \mathbf{\Theta}\big)\big\}, \mathbf{\Theta} \big)$. To tackle this issue, we propose an efficient algorithm by using the SSCA technique to find a suboptimal solution.\footnote{The effectiveness of SSCA technique in optimizing reflection coefficients has been validated in~\cite{Zhaomm2021TSS}.} Specifically, the IRS reflection coefficients are iteratively updated by maximizing the concave surrogate function of the objective function in Problem (\textbf{P5.1}), which is given by $\bar f^{(i)}(\mathbf{v})$ with $i$ representing the index of SSCA iteration.
	
	Considering that $\boldsymbol{\varsigma}$ is fixed when solving Problem (\textbf{P5.1}), we drop the argument $\boldsymbol{\varsigma}$ and denote $\big\{\mathbf{w}_{k}(\mathbf{H}(t),\mathbf{\Theta}) \big\}$ and $R_{k}(\mathbf{H}(t),\big\{\mathbf{w}_{k}(\mathbf{H}(t), \mathbf{\Theta})\big\}, \mathbf{\Theta}) $ as the short-timescale transmit beamforming vectors and the achievable rate of the $k$-th user for simplicity, respectively. First, $T_{\rm H}$ $ (T_{\rm H} \ll T) $ channel samples (i.e., $ \mathbf{H}^{(i)}(t_{\rm H})$) are generated based on S-CSI with $ t_{\rm h}\in\mathcal{T}_{\rm H}\triangleq\{1,\ldots,T_{\rm H}\} $, based on which, the concave surrogate function can be expressed as~\cite{Zhaomm2021TSS}
	\begin{align}\label{surrogate_fun}
		\bar f^{(i)}(\mathbf{v}) &= \sum_{k\in\mathcal{K}} \hat {R}_{k}^{(i)} + 
		2\mathcal{R}  \Big \{ (\mathbf{f}^{(i)})^H (\mathbf{v} -\mathbf{v}^{(i-1)} )  \Big\} \nonumber \\
		&-\tau ||\mathbf{v} -\mathbf{v}^{(i-1)}   ||^2,
	\end{align}
	where $\mathbf{v}^{(i-1)} $ denotes the IRS reflection coefficients obtained in the $(i-1)$-th iteration. Moreover,  $\hat {R}_{k}^{(i)}$ is the approximation of average achievable rate of the $k$-th user in the $i$-th SSCA iteration, which is updated as follows based on the $(i-1)$-th IRS reflection coefficients 
	\begin{align}
		\hat {R}_{k}^{(i)} &= (1-\rho^{(i)}) \hat {R}_{k}^{(i-1)}  \nonumber \\
		&+\frac{\rho^{(i)}}{T_{\rm H}} \sum_{t_{\rm H}\in\mathcal{T}_{\rm H} } R_{k}\big(\mathbf{H}^{(i)}(t_{\rm H}), \mathbf{w}_k^{(i)}(t_{\rm H}), \mathbf{v}^{(i-1)} \big), 
	\end{align}
	with $ \hat {R}_{k}^{(0)} = 0 $ and $\mathbf{w}_k^{(i)}(t_{\rm H}) \triangleq \mathbf{w}_{k}(\mathbf{H}^{(i)}(t_{\rm H}), \mathbf{v}^{(i-1)} )$ denoting the transmit beamforming vector corresponding to the $t_{\rm H}$-th generated channel sample $\mathbf{H}^{(i)}(t_{\rm H})$ under a fixed reflection coefficient vector $ \mathbf{v}^{(i-1)}$. The second term of~\eqref{surrogate_fun} represents an approximate partial derivative of average sum-rate, which is updated as
	\begin{align}
		\mathbf{f}^{(i)} &= (1-\rho^{(i)})\mathbf{f}^{(i-1)} \nonumber \\ 
		&+ \frac{\rho^{(i)}}{T_{\rm H}} \sum_{t_{\rm H}\in \mathcal{T}_{\rm H}} \mathbf{J}_{\mathbf{v}}\big(\mathbf{H}^{(i)}(t_{\rm H}), \mathbf{w}_k^{(i)}(t_{\rm H}), \mathbf{v}^{(i-1)} \big), 
	\end{align} 
	where  $\mathbf{J}_{\mathbf{v}}\big(\mathbf{H}^{(i)}(t_{\rm H}), \mathbf{w}_k^{(i)}(t_{\rm H}), \mathbf{v}^{(i-1)} \big)$
	is the Jacobian matrix of the achievable sum-rate $ R_{k}\big(\mathbf{H}^{(i)}(t_{\rm H}), \mathbf{w}_k^{(i)}(t_{\rm H}), \mathbf{v}^{(i-1)} \big) $  with respect to $\mathbf{v}$, the expression of which is provided in Appendix~\ref{Exp_Jv}.  The third term of~\eqref{surrogate_fun} is included to guarantee strong convexity, with $\tau$ being any positive constant.
	Therefore, by utilizing the randomly generated channel samples $\mathbf{H}^{(i)}(t_{\rm H})$ at the beginning of each SSCA iteration and their corresponding solutions $\mathbf{w}_k^{(i)}(t_{\rm H})$, the reflection coefficient vector $\mathbf{v}^{(i)}$ and the average sum-rate approximated by the concave surrogate function can be iteratively updated by solving the following  problem
	\begin{align}
		(\textbf{P5.2}):\; \max_{\mathbf{v}}&\quad \bar f^{(i)}(\mathbf{v}) \nonumber \\
		{\rm {s.t.}}&\quad \eqref{C_PS}. \nonumber
	\end{align}
	Note that (\textbf{P5.2}) is a convex optimization problem, for which its optimal solution, denoted by $ \mathbf{\bar v}^{(i)} \!=\! [\bar{v}_{1}^{(i)},\ldots,\bar{v}_{N}^{(i)}]^T$, is obtained below by using the Lagrange duality method with the details omitted due to limited space. 
	\begin{lemma}
		\rm
		The optimal solution to Problem (\textbf{P5.2}) is given~by
		\begin{equation}\label{re_coe_close}
			\bar{v}_{n}^{(i)} = \left\{
			\begin{aligned}
				&{v}_{n}^{(i-1)} + {f_n^{(i)}}/{\tau},&& \textrm{if}~|{v}_{n}^{(i-1)} + {f_n^{(i)}}/{\tau}|\le 1,\\
				& \frac{\tau {v}_{n}^{(i-1)} +  \frac{f_n^{(i)}}{\tau}}{\tau + \bar{\mu}^{\rm opt}}, &&\textrm{otherwise}, \\
			\end{aligned}
			\right.
		\end{equation}
		where $f_n^{(i)}$ denotes the $n$-th element of $ \mathbf{f}^{(i)} $ and $\bar{\mu}^{\rm opt}  = |\tau{v}_{n}^{(i-1)} + f_n^{(i)}|-\tau  $ denotes the optimal dual variable.
	\end{lemma}
	
	Based on~\eqref{re_coe_close}, the reflection coefficients in each SSCA iteration can be updated as
	\begin{align}
		\mathbf{v}^{(i)}  = (1-\delta^{(i)}) 	\mathbf{v}^{(i-1)}  + \delta^{(i)} \mathbf{\bar v}^{(i)}.
	\end{align}
	Herein, $\rho^{(i)}$ and $\delta^{(i)}$ are adjusted as follows for controlling the convergence speed of the SSCA method\footnote{The update rules of $ \mathbf{v}^{(i)} $ and  $\delta^{(i)}$ need to satisfy Assumption 5 in~\cite{Liu_CSSCA}.}
	\begin{align}
		\rho^{(i)} = \frac{1}{(1+i)^{0.8}},~\textrm{and}~
		\delta^{(i)} = \frac{2}{2+i}.
	\end{align}
	The above procedures are repeated until convergence of the inner problem is achieved; we thus obtain the average sum-rate (i.e., objective function of Problem (\textbf{P5.1})) and its corresponding reflection coefficients, denoted by $\bar R_{\rm sum}\big(\boldsymbol{\varsigma},\tilde{\mathbf{v}}(\boldsymbol{\varsigma})\big)$ and $\tilde{\mathbf{v}}$, respectively.

	\textbf{\underline{Outer problem}:} Given the inner problem, the outer problem is to optimize the antenna/surface configurations $\boldsymbol{\varsigma}$ for maximizing $\bar R_{\rm sum}\big(\boldsymbol{\varsigma},\tilde{\mathbf{v}}(\boldsymbol{\varsigma})\big)$, which can be formulated as
	\begin{align}
		(\textbf{P6}):\; \max_{\boldsymbol{\varsigma}}&\quad \bar R_{\rm sum}\big(\boldsymbol{\varsigma},\tilde{\mathbf{v}}(\boldsymbol{\varsigma})\big)  \nonumber  \\
		{\rm {s.t.}}&\quad\eqref{C_MA1},\eqref{C_MA2},\eqref{M_Rot},\eqref{C_Rot}. \nonumber
	\end{align}
	Considering that the optimization variables are highly coupled in the average sum-rate, we extend the DE algorithm in the outer problem to jointly update the rotation angles and positions of the 6DMA, as well as the rotation angle of the R-IRS.
	In the extended DE algorithm, the population of $P$ individuals in the $s$-th DE iteration is given by 
	\begin{align}
		\tilde{\mathcal{P}} ^{(s)}= \{  \boldsymbol{\varsigma}_{1}^{(s)},\boldsymbol{\varsigma}_{2}^{(s)},\ldots,\boldsymbol{\varsigma}_{P}^{(s)}\},~p\in\{1,2,\ldots,P\},
	\end{align}
	where $\boldsymbol{\varsigma}_{p}^{(s)} = \big[  \big( \mathbf{q}_{p}^{(s)}\big)^T, \big(\boldsymbol{\zeta}_{p}^{(s)} \big)^T \big]^T$ with $\boldsymbol{\zeta}_{p}^{(s)} = [\psi_{p}^{(s)}, \phi_{p}^{(s)}]^T$ denoting the rotation angles of the $p$-th individual in the $s$-th iteration. Then, the fitness function (i.e., the objective function of the outer problem) for the multi-user case can be expressed~as 
	\begin{align}\label{fit_MU}
		\mathcal{F}_{2} ( \boldsymbol{\varsigma}_{p}^{(s)} ) = \bar R_{\rm sum}\big(\boldsymbol{\varsigma}_{p}^{(s)},
		\mathbf{v}( \boldsymbol{\varsigma}_{p}^{(s)} )\big)
		\!-\! \eta \mathcal{B}_1(\mathbf{q}_{p}^{(s)} )  \big| \mathcal{B}_2(\mathbf{q}_{p}^{(s)} ) \big|. 
	\end{align}
	Note that the first term of~\eqref{fit_MU} is the average sum-rate obtained by solving the inner problem. Then, mutation, crossover, and selection operations are performed in the outer problem to update the individuals for finding the suboptimal rotation angle and positions of 6DMA, as well as rotation angle of IRS.\footnote{The mutation, crossover, and selection operations in the extended DE algorithm have no differences apart from the individuals $ \boldsymbol{\varsigma}_{p}^{(s)}$ and $\mathbf{q}_{p}^{(s)}$.} Since variables $\{\mathbf{q},\boldsymbol{\zeta}\}$ are jointly optimized via the extended DE algorithm, the following operation should be performed to satisfy constraints~\eqref{C_MA2} and ~\eqref{C_Rot}.
	\begin{equation}
		\big[\mathbf{\tilde u}_{p}^{(s)}\big]_{j}\!\!=\!\!\left\{
		\begin{aligned}
			&\max\big\{ \min\big\{\big[\tilde{\mathbf{ u}}_{p}^{(s)}\big]_{j}, q_{\rm max}\big\}, q_{\rm min} \big\}, 1\le j\le M,   \\
			&\max\big\{ \min\big\{\big[\tilde{\mathbf{ u}}_{p}^{(s)}\big]_{j}, \psi_{\rm max}\big\}, \psi_{\rm min} \big\}, j= M+1, \\
			&\max\big\{ \min\big\{\big[\tilde{\mathbf{ u}}_{p}^{(s)}\big]_{j}, \phi_{\rm max}\big\}, \phi_{\rm min} \big\}, j= M+2. \nonumber
		\end{aligned}
		\right.
	\end{equation}
	
	After $S$ iterations, the suboptimal solution for long-term optimization, denoted as $\boldsymbol{\varsigma}=\boldsymbol{\varsigma}_{\rm best}^{(S)}$, and the corresponding reflection coefficients can be obtained.
	\begin{remark}[Algorithm convergence and computational complexity]\rm 
		We first analyze the convergence of proposed algorithm, referred to as the DE-SSCA algorithm.
		Following the convergence proof of the SSCA technique in~\cite{Liu_CSSCA}, the convergence of the inner problem is guaranteed. For the extended DE algorithm that addresses the outer problem, the update of $\boldsymbol{\varsigma}_{\rm best}^{(s)}$ satisfies $ \mathcal{F}_{2} ( \boldsymbol{\varsigma}_{\rm best}^{(s)} ) \ge  \mathcal{F}_{2} ( \boldsymbol{\varsigma}_{\rm best}^{(s-1)} )$, which ensures the convergence of proposed DE-SSCA algorithm. The computational complexity of the proposed DE-SSCA algorithm is determined by the inner problem of optimizing the reflection coefficients of the R-IRS, which is in the order of $\mathcal{O}\big(I_S T_{\rm H} (I_{\mu} K M^3)\big)$ with $I_{\mu}$ and $I_S$ denoting the numbers of WMMSE and SSCA iterations, respectively. Consequently, the overall computational complexity of the proposed DE-SSCA algorithm is in the order of  $ \mathcal{O}\big( S P I_S T_{\rm H} ( I_{\mu} K M^3)\big) $.
	\end{remark}

	\subsection{Low-Complexity Algorithm}
	While the proposed DE-SSCA algorithm
	offers a high-quality solution to Problem (\textbf{P1}), its computational complexity is significantly high, and thus it may be unaffordable in practice. To address this issue, we propose a low-complexity algorithm that reduces computational complexity while achieving satisfactory performance. 
	
	Inspired by the maximization of expected equivalent channel power gain in the single-user scenario, the key idea of the proposed low-complexity algorithm is to optimize the reflection coefficients of the R-IRS for maximizing the average sum-channel-gain (SCG) based on S-CSI.
	As such, we define the instantaneous SCG as
	\begin{align}
		G_{\rm sum} =  \sum_{k\in\mathcal{K}} \big\| \mathbf{h}_{k}(\mathbf{q})+\mathbf{g}_{k}(\mathbf{q},\boldsymbol{\zeta},\mathbf{\Theta}) \big\|_2^2.
	\end{align}
	Accordingly, the average SCG based on S-CSI is given by
	\begin{align}
		G_{\rm avg}(\mathbf{q},\boldsymbol{\zeta}) &= \mathbb{E}  \big[ G_{\rm sum} \big] 
		\overset{(h)}{=} \sum_{k\in\mathcal{K}} \text{Tr}(\mathbf{H}_{{\rm eff},k} \mathbf{V}),
	\end{align}
	where $(h)$ holds due to {\bf Lemma~\ref{closed_form1}}. Therefore, for any feasible $\mathbf{q}$ and $\boldsymbol{\zeta}$, the inner problem (\textbf{P5.1}) is reformulated to optimize the reflection coefficients of the R-IRS for maximizing the SCG. To this end, the corresponding optimization problem can be formulated as
	\begin{align}
		(\textbf{P5.3}):\; \max_{\mathbf{V}}&\quad \text{Tr}(   \mathbf{H}_{{\rm eff}} \mathbf{V})  \nonumber  \\
		{\rm {s.t.}}&\quad \eqref{V_1}-\eqref{V_3}, \nonumber
	\end{align} 
	where  $\mathbf{H}_{{\rm eff}} =  \sum_{k\in\mathcal{K}} \mathbf{H}_{{\rm eff},k}$. Similar to Problem (\textbf{P3.2}), this problem can be solved via CVX tool after relaxing the rank-one constraint in~\eqref{V_3}. After obtaining the reflection coefficient vector $\mathbf{v}_{\rm Low}$, the average sum-rate can be obtained~as
	\begin{align}
		\frac{1}{T_{\rm H}}\sum_{t_{\rm H}\in\mathcal{T}_{\rm H}} R_{k}\big(\mathbf{H}(t_{\rm H}),\mathbf{w}_{k}(\mathbf{H}(t_{\rm H}),\mathbf{\Theta}_{\rm Low}),\mathbf{\Theta}_{\rm Low}\big), 
	\end{align}
	and the fitness function can be correspondingly updated. Afterwards, the extended DE method in Section~\ref{Sec:IV-A-2} can be used to optimize $\mathbf{q}$ and $\boldsymbol{\zeta}$, with the details omitted for brevity. 
	
	\begin{remark}[Convergence and computational complexity of proposed low-complexity algorithm]\rm 
		Since the convergence of the extended DE algorithm has been established, the convergence of proposed low-complexity algorithm is also guaranteed. For the computational complexity, the complexity of the proposed low-complexity algorithm is in the order of
		$\mathcal{O}\big( S P  T_{\rm H} ( I_{\mu} K M^3) \big)$. Compared with DE-SSCA algorithm in Section~\ref{Sec:IV},
		the proposed low-complexity algorithm avoids the iterative optimization of the R-IRS reflection coefficients (which typically requires over 100 iterations, as shown in Section~\ref{Sec:SR}), thereby achieving much lower computational complexity. 
	\end{remark}
	
	\vspace{-16pt}
	\section{Numerical Results}\label{Sec:SR}
	In this section, we present numerical results to validate the convergence and efficacy of the proposed algorithms. The short-timescale complex path-response coefficients follow the distance-dependent path loss model, which is expressed as $ \beta  = \frac{\lambda}{4\pi r}$, where $r$ represents the distance between two nodes. 
	The 6DMA-BS and R-IRS are deployed at $(1,1,0)$ m and $(0,0,0)$~m, respectively, while the users are randomly distributed within a circular region centered at $(4,-18,0)$ m with a radius of $3$ m. 
	Moreover, the AoAs and AoDs of each channel follows a uniform distribution over $[\frac{\pi}{6},\frac{5\pi}{6}]$, and the movement region and rotation region are set as $\mathcal{C}_{ \mathbf{q}} =[-3\frac{(M-1)d}{2},3\frac{(M-1)d}{2}]  $, with the maximum aperture of MA array being $D = 3D_{\rm ULA}$, where $D_{\rm ULA} = (M-1)d$ represents the aperture of uniform LA (ULA), $ \mathcal{C}_{\psi} = [-\frac{\pi}{6},\frac{\pi}{6}] $ and  $\mathcal{C}_{\phi} = [-\frac{\pi}{6},\frac{\pi}{6}]$, respectively. 
	Unless otherwise specified, the other parameters are presented as follows: $ M =10 $, $N = 20\times 10 = 200$, $f_c = 6$ GHz,  $ L = L_{{\rm r},k} =L_{{\rm t},k} =  5 $, $K = 4$, $P_{\rm t} = 30$~dBm,  $\sigma^2 = -40$ dBm, $P = 50$, $S = 50$, $F = 0.6$, $C_{\rm R}=0.9$, $\eta = 1000$, $T_{\rm H} =50$, and $\tau =0.015$.

	For performance comparison, we consider the following benchmark schemes.
	\begin{itemize}
	\item \emph{Fixed-configuration scheme}:
	In this scheme, all antenna and surface configurations are fixed, while the reflection coefficients of IRS and the transmit beamforming are optimized using the proposed algorithm.
	\item \emph{6DMA + F-IRS (F-IRS refers to fixed-rotation IRS) scheme}:
	In this scheme, the rotation angle of R-IRS is fixed, while its reflection coefficients, 6DMA positions and rotation, as well as the transmit beamforming, are optimized using the proposed algorithm.
	\item \emph{R-IRS only scheme}: 
	In this scheme, the positions and rotation of the BS ULA are fixed, while the rotation angle and reflection coefficients of R-IRS, as well as the transmit beamforming, are optimized using the proposed algorithm.
	\item \emph{Rotatable 6DMA + F-IRS scheme}:
	Similar to the 6DMA scheme, but only the rotation angle of 6DMA is optimized, with its positions remaining fixed.
	 
	\item \emph{Positionable 6DMA + F-IRS scheme}:
	Similar to the 6DMA scheme, but only the 6DMA positions are optimized, with its rotation angle fixed.
	\end{itemize}
	\subsection{Single-user Scenario}
	We present numerical results for the single-user scenario to demonstrate the effectiveness of the proposed algorithm and the superiority of the proposed joint R-IRS and 6DMA-BS architecture in improving average sum-rate.
	
	\begin{figure}[t]
		\centering
		\includegraphics[width=0.35\textwidth]{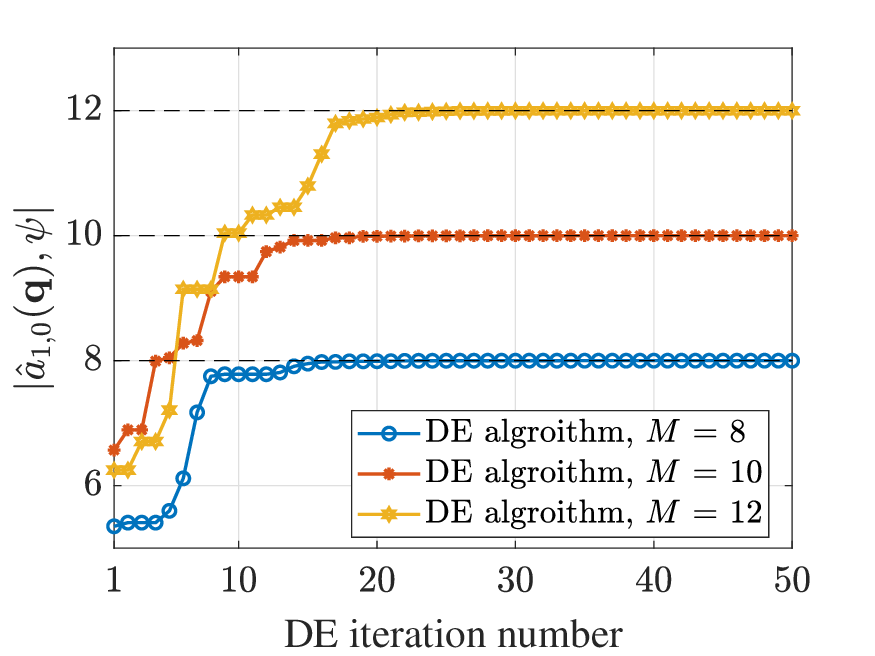}
		\caption{$\left|\hat{a}_{1,0}(\mathbf{q},\psi)\right|$ versus DE iteration number.} 
		\label{Fig:Convergence_SU1}
		\vspace{-16pt}
	\end{figure}
	\subsubsection{Convergence of the Proposed DE Algorithm} 
	In Fig.~\ref{Fig:Convergence_SU1}, the convergence of proposed DE algorithm for solving Problem (\textbf{P3.1}) under different numbers of MAs is presented. It is observed that the proposed DE algorithm converges to a near-optimal performance value (i.e., maximum value $M$)  after about 30 iterations in the single-user scenario. This demonstrates that the proposed DE algorithm is capable of obtaining high-quality solutions to Problem (\textbf{P3.1}).
	
	\subsubsection{Effect of NLoS Paths} 
	\begin{figure}[t]
		\centering
		\includegraphics[width=0.35\textwidth]{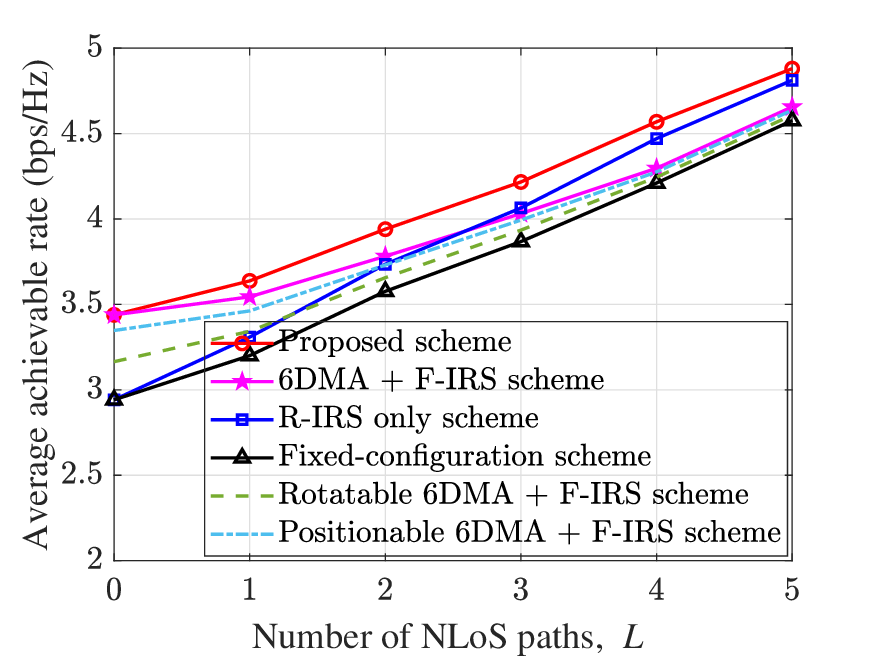}
		\caption{Average sum-rate versus number of NLoS paths.} 
		\label{Fig:SUpaths}
		\vspace{-16pt}
	\end{figure}
	In Fig.\ref{Fig:SUpaths}, the average rate versus the number of NLoS paths under different schemes is presented. Several observations are made as follows. First, the average rate of all schemes monotonically increases with the number of NLoS paths, since more NLoS paths provide higher effective path gains. Second, the proposed scheme consistently outperforms the benchmarks, achieving the highest average sum-rate and thereby validating its effectiveness.
	Moreover, Fig.~\ref{Fig:SUpaths} shows that when $L = 0$ (i.e., the LoS scenario), the R-IRS does not bring any improvement in the user rate performance. In this case, the rate enhancement of the proposed scheme is solely attributed to the deployment of 6DMA, where both rotation and position optimization contribute to the performance gain, as discussed in {\bf Remark~\ref{Rem2}}. By contrast, when $L > 0$ (i.e., the multi-path scenario), the rotation of the R-IRS introduces additional DoFs that enhance the reflected link gain, leading to a significant rate improvement.

	\subsubsection{Effect of Movement Region}
	\begin{figure}[t]
		\centering
		\includegraphics[width=0.35\textwidth]{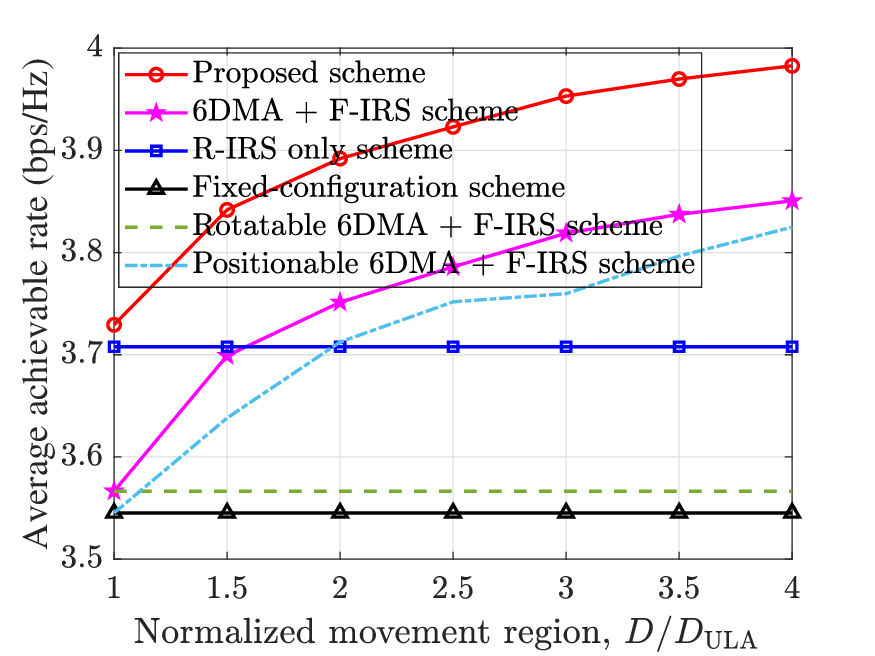}
		\caption{Average sum-rate versus movement region.} 
		\label{Fig:SUaperture}
		\vspace{-16pt}
	\end{figure}
	In addition, we present the impact of the allowable movement region on the average user rate across all schemes with $L=2$. As the movement region expands from $D_{\rm ULA}$ to $4D_{\rm ULA}$, the proposed scheme, the 6DMA + F-IRS scheme, and the positionable 6DMA + F-IRS scheme achieve increased average user rates, while the performance of other schemes remains unchanged. This performance gain stems from the enhanced positionable DoFs enabled by a larger movement region, which increases the value of $\left|\hat{a}_{1,0}(\mathbf{q},\psi)\right|$ and thereby improves the average user rate. 
	Furthermore, by introducing the rotatable DoFs at both the BS and IRS, the incorporation of rotation leads to further rate improvement, especially for the R-IRS only scheme, which exploits rotation for multi-path alignment. Consequently, the proposed scheme demonstrates significant rate gains by integrating both 6DMA and R-IRS.


	\subsection{Multi-user Scenario}
	\subsubsection{Convergence of the Proposed Algorithms}
	\begin{figure}[t]
		\centering
		\includegraphics[width=0.35\textwidth]{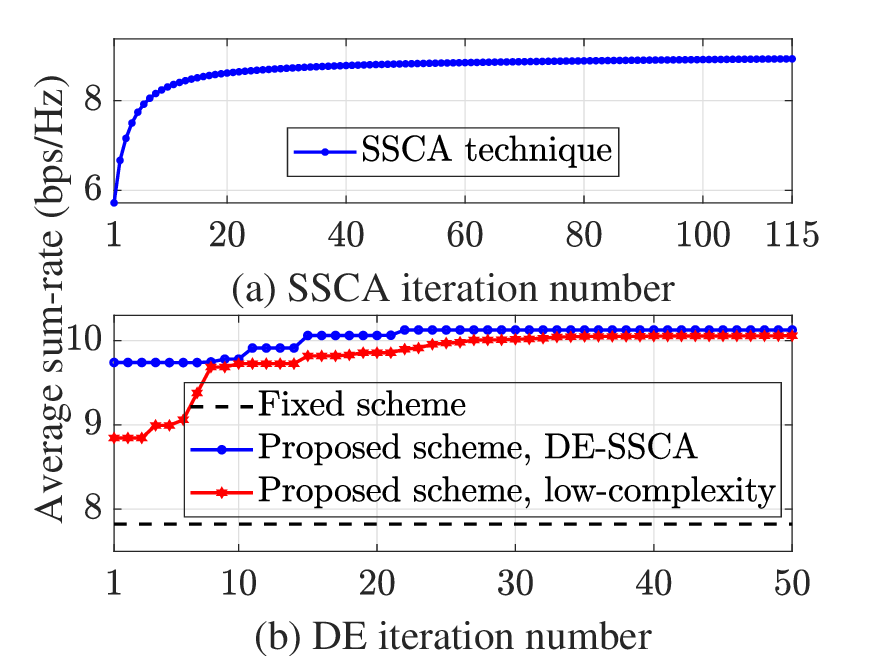}
		\caption{Convergence of SSCA technique and proposed two algorithms.} 
		\label{Fig:MUconvergence}
		\vspace{-16pt}
	\end{figure}
	We first show in Fig.~\ref{Fig:MUconvergence} the convergence performance for the proposed two algorithms. 
	In Fig.~\ref{Fig:MUconvergence}(a), we plot the value of the concave surrogate function in~\eqref{surrogate_fun} versus the number of SSCA iterations. It is observed that the average sum-rate converges to a stationary point after about 100 iterations.
	In addition, Fig.~\ref{Fig:MUconvergence}(b) shows convergence performance for the proposed two algorithms. It is observed that the average sum-rate of these two algorithms converges to a stationary point after 40 iterations.
	Compared with the fixed scheme, both the proposed DE-SSCA and the low-complexity algorithms yield a substantial improvement in average sum-rate, thereby validating the effectiveness of the extended DE algorithm in jointly optimizing the 6DMA positions and rotation angles, as well as the R-IRS rotation angle. Although the DE-SSCA algorithm achieves the highest average sum-rate, it requires more than 100 iterations for optimizing the reflection coefficients, which may be computationally prohibitive. By contrast, the proposed low-complexity algorithm achieves an average sum-rate comparable to that of the DE-SSCA algorithm, but avoids iterative optimization. Given the prohibitive computational complexity of DE-SSCA algorithm, we focus on evaluating the performance of proposed scheme for the multi-user scenario using the low-complexity algorithm.

	
	\subsubsection{Effect of Transmit Power}
	\begin{figure}[t]
		\centering
		\includegraphics[width=0.35\textwidth]{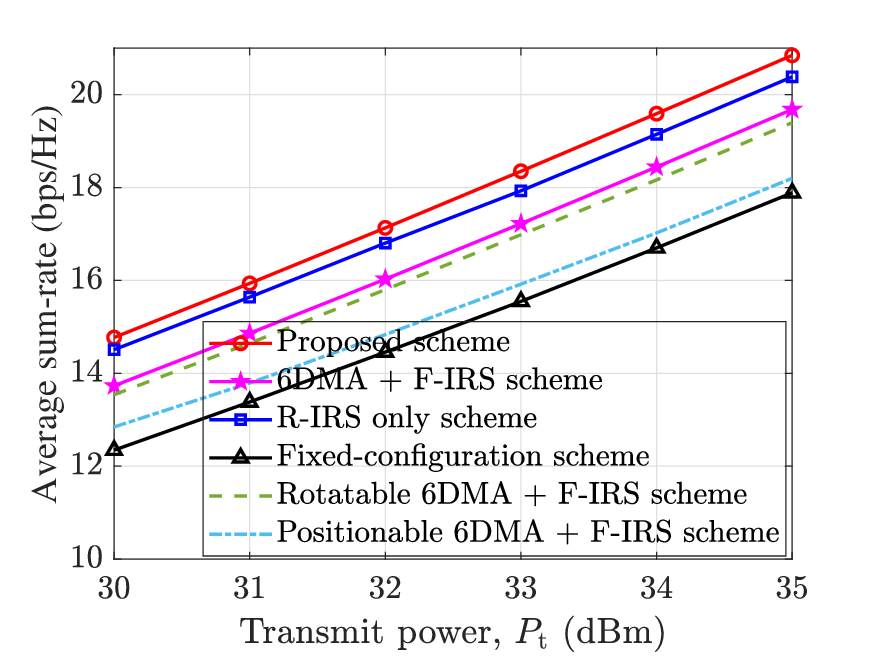}
		\caption{Average sum-rate versus transmit power.} 
		\label{Fig:RateversusPt}
		\vspace{-16pt}
	\end{figure}
	In addition, Fig.~\ref{Fig:RateversusPt} plots the average sum-rate versus the transmit power for all schemes with the number of reflecting elements being $100$.
	Among them, the proposed scheme achieves the highest sum-rate, owing to the joint deployment of R-IRS and 6DMA, which provides additional spatial DoFs to enhance rate performance. Moreover, both the R-IRS only and 6DMA + F-IRS schemes significantly outperform the fixed scheme. The gain of the former confirms that IRS rotation improves the average rate performance, while the gain of the latter is attributed to optimizing the position and rotation of 6DMA. 
	
	\subsubsection{Effect of Reflecting Elements}
	\begin{figure}[t]
		\centering
		\includegraphics[width=0.35\textwidth]{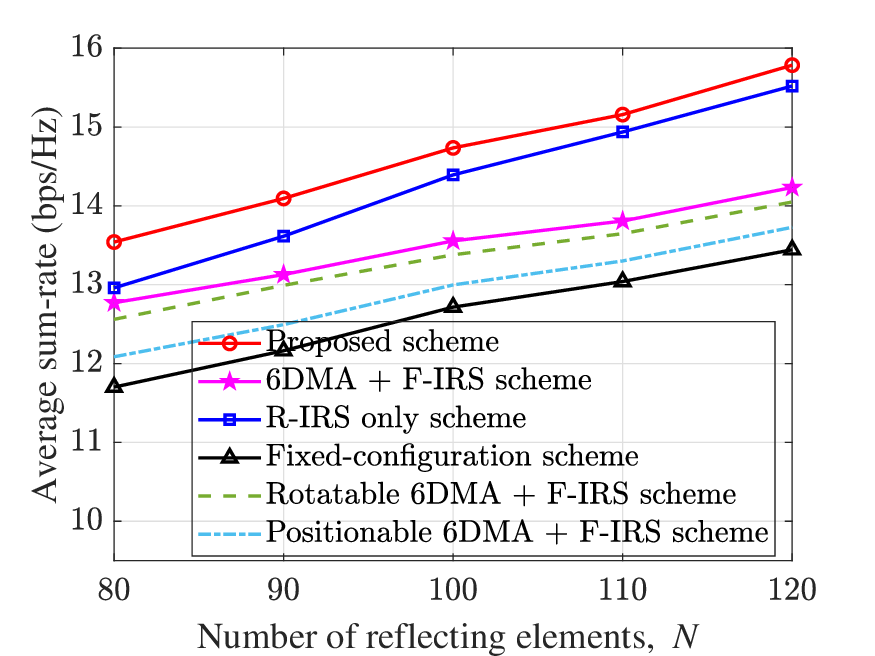}
		\caption{Average sum-rate versus number of reflecting elements.} 
		\label{Fig:Rateversusele}
		\vspace{-16pt}
	\end{figure}
	Last, in Fig.~\ref{Fig:Rateversusele}, we present the  average sum-rate versus the number of reflecting elements $N$ under different schemes. 
	As expected, the rate performance of all schemes improves with increasing $N$, since more reflecting elements provide additional reliable reflected links for information transmission.
	An interesting observation is that the performance gap between R-IRS only and 6DMA + F-IRS schemes gradually increases as $N$ grows, leading to a significant performance gain for the proposed scheme. This improvement stems from the rotatable DoFs of R-IRS, which substantially strengthen the reflected links and thereby boost the sum-rate.

	\section{Conclusions}\label{Sec:Con}
	In this paper, we studied the performance gains of  multi-functional antenna/surface-assisted multi-user communication systems.
	To maximize the average sum-rate under a practical TTS transmission protocol, we first considered the single-user scenario, where the original problem was transformed into maximizing the expected equivalent channel power gain. Based on this formulation, we revealed that: 1) a sparse-array configuration of 6DMA-BS enables efficient coordination between the direct and reflected links, and 2) the rotatable DoFs of R-IRS facilitate effective multi-path alignment, both of which contribute to an enhanced equivalent channel gain. 
	For the multi-user case, we developed the DE-SSCA algorithm to tackle the resulting highly coupled optimization problem. Furthermore, a low-complexity algorithm was proposed to
	reduce computational complexity. Numerical results showed that the proposed scheme substantially outperforms various benchmark schemes, highlighting the significant performance improvements achieved by jointly exploiting the spatial DoFs of 6DMA-BS and R-IRS. 
	\begin{appendices}
		\vspace{-10pt}
		\section{}\label{App1}
		As given by~\eqref{derivation}, the objective function consists of three parts. Based on the statistical information of channels, the first part can be reformulated as~\eqref{Expection1}, as shown at the top of next page. 
		\setcounter{equation}{\value{equation}}
		\begin{figure*}[ht]
			\begin{align}\label{Expection1}
				&\mathbb{E} \big[ \mathbf{h}^H_{1}(\mathbf{q},\psi )\mathbf{h}_{1}(\mathbf{q},\psi)\big]  
				=\mathbb{E} \big[ \big( \mathbf{A}_{{\rm t},1}  ( \mathbf{q},\psi)\boldsymbol{\beta}_{{\rm t},1}\big) ^H \mathbf{A}_{{\rm t},1} ( \mathbf{q},\psi) \boldsymbol{\beta}_{{\rm t},1}\big] 
				= \mathbb{E} \big[  \boldsymbol{\beta}^H_{{\rm t},1} \mathbf{A}^H_{{\rm t},1} (\mathbf{q},\psi) \mathbf{A}_{{\rm t},1}  (\mathbf{q},\psi) \boldsymbol{\beta}_{{\rm t},1}
				\big] \nonumber \\
				\overset{(a)}{=}  &\Big(
				|\tilde{\beta}_{1,0}|^2 \tilde{\mathbf{a}}^H_{{1},{0}} (\mathbf{q},\psi) \tilde{\mathbf{a}}_{{1},{0}} (\mathbf{q},\psi ) 
				+\sum_{\ell_{{\rm t},1}=1}^{L_{{\rm t},1}}  \tilde{\sigma}_{1,\ell_{{\rm t},1}}^{2} \tilde{\mathbf{a}}^H_{1,{\ell_{{\rm t},1}}} (\mathbf{q},\psi ) \tilde{\mathbf{a}}_{1,{\ell_{{\rm t},1}}} (\mathbf{q},\psi)
				\Big) 
				\overset{(b)}{=} M \Big(|\tilde{\beta}_{1,0}|^2 + \sum_{\ell_{{\rm t},1}=1}^{L_{{\rm t},1}}  \tilde{\sigma}_{1,\ell_{{\rm t},1}}^{2}   \Big) \triangleq c_1, 
			\end{align}
			\vspace{-4pt}
			\hrulefill
			\vspace{-12pt}
		\end{figure*}
		Specifically,  $(a)$ holds due to the fact that $\tilde{\beta}_{{1},\ell_{{\rm t},1}} \sim \mathcal{CN}(0, \tilde{\sigma}_{1,\ell_{{\rm t},1}}^{2} ) $ with $\ell_{{\rm t},1} \in \{1,2,\ldots,L_{{\rm t},1}\}$ and $(b)$ holds  due to the expression~\eqref{atk}, which implies that  $\tilde{\mathbf{a}}^H_{{1},{\ell_{{\rm t},1}}} (\mathbf{q},\psi) \tilde{\mathbf{a}}_{{1},{\ell_{{\rm t},1}}} (\mathbf{q},\psi) = M $. 
		Let denote $ \bar{\mathbf{G}}(\phi)  =\mathbf{G}_{\rm r}(\phi)  \mathbf{\Xi} \mathbf{G}_{\rm r}^H(\phi) $ with 
		\begin{align}
			\mathbf{\Xi} = M \diag\big(|\beta_0|^2,\sigma_{1}^2,\ldots,\sigma_{L}^2\big).
		\end{align}
		
		\setcounter{equation}{\value{equation}}
		\begin{figure*}[ht]
			\begin{align}\label{Expection2}
				&~~~~\mathbb{E} \big[ \mathbf{g}^H_{1}(\mathbf{q},\boldsymbol{\zeta},\mathbf{\Theta} ) \mathbf{g}_{1}(\mathbf{q},\boldsymbol{\zeta},\mathbf{\Theta} ) \big] 
				= \mathbb{E} 
				\big[ \mathbf{v}^T 
				\text{diag}\big( 
				( \mathbf{A}_{{\rm r},1}  (\phi) \boldsymbol{\beta}_{{\rm r},1} )^H\big) 
				\mathbf{G}_{\rm r}(\phi) 
				\mathbf{\Sigma} 
				\mathbf{G}_{\rm t}^{H}(\mathbf{q},\psi)  
				\mathbf{G}_{\rm t}(\mathbf{q},\psi) \mathbf{\Sigma}^H  \mathbf{G}_{\rm r}^H(\phi) 
				\text{diag}(  \mathbf{A}_{{\rm r},1} (\phi )\boldsymbol{\beta}_{{\rm r},1} ) 
				\mathbf{v}^\dagger 
				\big]  \nonumber \\
				& \overset{(c)}{=}  \mathbf{v}^T\!\Big(
				|\bar{\beta}_{{1},0}|^2  
				\text{diag}\big(  \bar{\mathbf{a}}^H_{{1},{0}} (\phi ) \big) 
				\bar{\mathbf{G}}(\phi) 
				\text{diag}\big(  \bar{\mathbf{a}}_{{1},{0}} (\phi) \big) \!+\! 
				\sum_{\ell_{{\rm r},1}=1}^{L_{{\rm r},1}}  \bar{\sigma}_{1,\ell_{{\rm r},1}}^{2} \text{diag}\big( \bar{\mathbf{a}}^H_{{1},{\ell_{{\rm r},1}}} (\phi ) \big) 
				\bar{\mathbf{G}}(\phi) 
				\text{diag}\big( \bar{\mathbf{a}}_{{1},{\ell_{{\rm r},1}}} (\phi ) \big)
				\Big)\!
				\mathbf{v}^\dagger 
				\triangleq  \mathbf{v}^T \hat{\mathbf{G}}_{1} (\phi)  \mathbf{v}^\dagger,
			\end{align}
			\hrulefill
			\vspace{-12pt}
		\end{figure*}
		\setcounter{equation}{\value{equation}}
		\begin{figure*}[ht]
			\begin{align}\label{Expection3}
				&~~~~\mathbb{E} \big[2\mathcal{R} \big\{ \mathbf{h}^H_{1}(\mathbf{q},\psi) \mathbf{g}_{1}(\mathbf{q},\boldsymbol{\zeta},\mathbf{\Theta}) \big\}\big] 
				= \mathbb{E} \big[ 2\mathcal{R} \big\{
				\big(  \mathbf{r}_{1}^H (\phi )  \mathbf{\Theta} \mathbf{G}( \mathbf{q},\boldsymbol{\zeta}) \big) 
				\mathbf{A}_{{\rm t},1}  ( \mathbf{q},\psi )\boldsymbol{\beta}_{{\rm t},1}
				\big\}
				\big] \nonumber \\
				& = \mathbb{E} \big[ 2\mathcal{R} \big\{
				\mathbf{v}^T 
				\text{diag}\big( ( \mathbf{A}_{{\rm r},1}  (\phi) \boldsymbol{\beta}_{{\rm r},1})^H \big) 
				\mathbf{G}_{\rm r}(\phi) 
				\mathbf{\Sigma} 
				\mathbf{G}_{\rm t}^{H}(\mathbf{q},\psi) 
				\mathbf{A}_{{\rm t},1}  ( \mathbf{q},\psi )\boldsymbol{\beta}_{{\rm t},1}
				\big\}
				\big] \nonumber \\
				&\overset{(d)}{=} 2\mathcal{R} \big\{ \omega_{1} \mathbf{v}^T 
				\text{diag}\big(  \bar{\mathbf{a}}^H_{{1},{0}} (\phi) \big)  
				\mathbf{a}_{{\rm r},0} ( \phi )
				\mathbf{a}_{{\rm t},0}^H ( \mathbf{q},\psi )
				\tilde{\mathbf{a}}_{{1},{0}} (\mathbf{q},\psi )
				\big\} 
				\triangleq 2\mathcal{R} \big\{ \omega_{1} \mathbf{v}^T   \hat{\mathbf{a}}_{{1},{0}} (\phi )  
				\hat{a}_{{1},{0}}(\mathbf{q},\psi)
				\big\}, 
			\end{align}
			\vspace{-16pt}
			\hrulefill
		\end{figure*}
		
		Similarly, the second part can be reformulated as~\eqref{Expection2} with $\hat{\mathbf{G}}_{1} (\phi) =
		\Big(
		|\bar{\beta}_{{1},0}|^2  
		\text{diag}\big(  \bar{\mathbf{a}}^H_{{1},{0}} (\phi ) \big) 
		\bar{\mathbf{G}}(\phi) 
		\text{diag}\big(  \bar{\mathbf{a}}_{{1},{0}} (\phi)  \big) 
		+  \sum_{\ell_{{\rm r},1}=1}^{L_{{\rm r},1}}  \bar{\sigma}_{1,\ell_{{\rm r},1}}^{2} \text{diag}\big( \bar{\mathbf{a}}_{{1},{\ell_{{\rm r},1}}}^H (\phi) \big) 
		\bar{\mathbf{G}}(\phi) 
		\text{diag}\big(\bar{\mathbf{a}}_{{1},{\ell_{{\rm r},1}}} (\phi)  \big)
		\Big)$, and $(c)$ holds  due to the fact that $\beta_{\ell} \sim \mathcal{CN}\left(0, \sigma_{\ell}^2 \right),\forall \ell \in \{1,2,\ldots, L\}  $ and $\bar{\beta}_{{1},\ell_{{\rm r},1}} \sim \mathcal{CN}(0, \bar{\sigma}_{1,\ell_{{\rm r},1}}^{2} ),\forall \ell_{{\rm r},1} \in \{1,2,\ldots,L_{{\rm r},1}\}$, respectively.
		Additionally, the third part can be reformulated as~\eqref{Expection3}, where 
		$\hat{\mathbf{a}}_{{1},{0}} (\phi) = \text{diag}(  \bar{\mathbf{a}}^H_{{1},{0}} \left(\phi \right)) \mathbf{a}_{{\rm r},0} \left( \phi \right)$,
		$\hat{a}_{{1},{0}}\left(\mathbf{q},\psi \right)  =\mathbf{a}_{{\rm t},0}^H \left( \mathbf{q},\psi\right)
		\tilde{\mathbf{a}}_{{1},{0}} \left(\mathbf{q},\psi \right)$, and 
		$\omega_{1} =\beta_{0} \bar{\beta}_{{1},0}  \tilde{\beta}_{{1},0} $. Thus, the proof of {\bf Lemma~\ref{closed_form1}} is completed.
			
		\vspace{-10pt}
		\section{}\label{Decoupl_app}
		The last  two terms  in~\eqref{obj_closed1} can be rewritten as
		\begin{align}
			& \mathbf{v}^T \hat{\mathbf{G}}_{1}  \left(\phi \right)  \mathbf{v}^\dagger + 2\mathcal{R} \big\{ \omega_{1} \mathbf{v}^T  \hat{\mathbf{a}}_{{1},{0}} \left(\phi \right)
			\hat{a}_{{1},{0}}\left(\mathbf{q},\psi \right)\big\}  \nonumber \\
			\overset{(e)}{=} &    \hat{\mathbf{v}}^T \hat{\mathbf{G}}_{1}  \left(\phi \right)  \hat{\mathbf{v}}^\dagger  + 2\mathcal{R} \big\{ \omega_{1} \hat{\mathbf{v}}^T  \hat{\mathbf{a}}_{{1},{0}} \left(\phi \right)
			\hat{a}_{{1},{0}}\left(\mathbf{q},\psi \right) e^{j 2 \pi \hat \theta} \big\}   \nonumber \\
			\overset{(f)}{=} &  \hat{\mathbf{v}}^T \hat{\mathbf{G}}_{1}  \left(\phi \right)  \hat{\mathbf{v}}^\dagger +  2 |\omega_{1}|  |\hat{a}_{{1},{0}}\left(\mathbf{q},\psi\right)|  \mathcal{R}  \big\{  \hat{\mathbf{v}}^T   \hat{\mathbf{a}}_{{1},{0}} \left(\phi \right) \big\}, 
		\end{align}
		where $ \hat {\mathbf{v}}=  e^{-j  \hat \theta} \mathbf{v} $ with $e^{j \hat \theta}$ denoting the  auxiliary phase shift, which is employed to compensate for the phase difference.  Note that (e) holds because of the equality $\mathbf{v}^T \hat{\mathbf{G}}_{1}  \left(\phi \right)   \mathbf{v}^\dagger = e^{j \hat \theta}  \hat{\mathbf{v}}^T \hat{\mathbf{G}}_{1}  \left(\phi \right)  \hat{\mathbf{v}}^\dagger e^{-j \hat \theta}  = \hat{\mathbf{v}}^T \hat{\mathbf{G}}_{1}  \left(\phi \right)  \hat{\mathbf{v}}^\dagger $ and (f) holds as a result of utilizing $e^{j \hat \theta}$  to compensate the phases of the $\omega_{1}$ and $ \hat{a}_{{1},{0}}\left(\mathbf{q},\psi \right) $. Thus, the proof of {\bf Lemma~\ref{decoupled}} is completed by substituting $\hat {\mathbf{v}}$ with ${\mathbf{v}}$.
		
		\section{}\label{Exp_Jv}
		For a given channel sample $\mathbf{H} \triangleq\{\mathbf{h}_{k},\mathbf{r}_{k},\mathbf{G} \}$, transmit beamforming vector $\mathbf{w}_{k}$, and reflection coefficient vector $\mathbf{v}$,  the general Jacobian matrix can be expressed as
		\begin{align}\label{J_matrix}
			\mathbf{J}_{\mathbf{v}}\left(\mathbf{H}, \mathbf{w}_k, \mathbf{v} \right) 
			= \sum_{k\in\mathcal{K}}\left(\frac{\mathbf{\Lambda}_{k}}{\Gamma_{k}}  - \frac{\mathbf{\Lambda}_{-k}}{\Gamma_{-k}}\right), 
		\end{align}
		where
$ \Gamma_{k} = \sum_{j=1}^{K}\left| (\mathbf{v}^T \text{diag}(\mathbf{r}_k^H) \mathbf{G} +\mathbf{h}_{k}^H)\mathbf{w}_{j} \right|^2 + \sigma^2$,
$	\Gamma_{-k} = \sum_{j=1,k\neq k}^{K}\left| (\mathbf{v}^T \text{diag}(\mathbf{r}_k^H) \mathbf{G} +\mathbf{h}_{k}^H)\mathbf{w}_{j} \right|^2 + \sigma^2$, 
{\small$ \mathbf{\Lambda}_{k} =\sum_{j=1}^{K}\big( \text{diag}(\mathbf{r}_{k}^H)\mathbf{G} \mathbf{w}_{j} \mathbf{w}_{j}^H \mathbf{G}^H \text{diag}(\mathbf{r}_{k}) \mathbf{v}^{\dagger}   + \diag(\mathbf{r}_{k}^H ) \mathbf{G}   \mathbf{w}_{j} \mathbf{w}_{j}^H \mathbf{h}_{k} \big)$}, 
and 
{\small$\mathbf{\Lambda}_{-k}  =  \sum_{j=1,j\neq k}^{K}\big( \text{diag}(\mathbf{r}_{k}^H)\mathbf{G} \mathbf{w}_{j} \mathbf{w}_{j}^H \mathbf{G}^H \text{diag}(\mathbf{r}_{k}) \mathbf{v} + \diag(\mathbf{r}_{k}^H ) \mathbf{G}   \mathbf{w}_{j} \mathbf{w}_{j}^H \mathbf{h}_{k} \big)$}. By substituting  $ \mathbf{H}^{(i)}(t_{\rm H})$, $\mathbf{w}_k^{(i)}(t_{\rm H})$, and $\mathbf{v}^{(i-1)} $ into~\eqref{J_matrix}, we thus obtain the Jacobian matrix.
		
	\end{appendices}
	
	\bibliographystyle{IEEEtran}
	\bibliography{Ref_MF_BSandIRS.bib}
	
\end{document}